\documentclass[10pt,journal,cspaper,compsoc]{IEEEtran}
\pdfoutput=1
 \usepackage{etex}

\usepackage[T1]{fontenc} 
\usepackage[utf8]{inputenc} 
\usepackage[linesnumbered]{algorithm2e}
\usepackage{multirow}
\usepackage{xcolor}
\usepackage{balance}
\usepackage{xspace}
\usepackage{listings}
\usepackage{comment}
\usepackage{mdframed}
\usepackage{framed}
\usepackage{amsmath}
\usepackage{cases}
\usepackage{float}

\usepackage[T1]{fontenc} 
\usepackage[utf8]{inputenc} 
\usepackage[colorlinks,citecolor=blue,urlcolor=blue,linkcolor=blue]{hyperref}
\usepackage{tabularx}
\usepackage{textcomp}
\usepackage{listings}
\usepackage{xspace}
\usepackage{graphicx}
\usepackage{xcolor}
\usepackage{acronym}

\newcounter{rqs}
\newcommand{\nopol}{{\sc Nopol}\xspace}

\def\ourif{{\sc if}\xspace}
\newcommand{\tabincell}[2]{\begin{tabular}{@{}#1@{}}#2\end{tabular}} 
\newcommand{\tabfootnote}[1]{\raggedright \scriptsize #1}

\newcommand{\buggycondition}{buggy \ourif condition\xspace}
\newcommand{\precondition}{missing precondition\xspace}
\newcommand{\buggyconditions}{buggy \ourif conditions\xspace}
\newcommand{\preconditions}{missing preconditions\xspace}

\newcommand{\buggyandpres}{\buggyconditions and \preconditions}

\newcommand{\mycode}[1]{{\small \texttt{#1}}\xspace}

\newcommand{\mytrue}[0]{\mycode{true}}
\newcommand{\myfalse}[0]{\mycode{false}}

\newcommand{\numbug}{22\xspace}

\begin{document}

\title{Nopol: Automatic Repair of Conditional Statement Bugs in Java Programs}

\author{Jifeng~Xuan,~\IEEEmembership{Member,~IEEE,} 
        Matias~Martinez,
        Favio~DeMarco$^{\dag}$,
        Maxime~Cl\'{e}ment$^{\dag}$,
        Sebastian~Lamelas~Marcote$^{\dag}$, 
        Thomas~Durieux$^{\dag}$, 
        Daniel~Le~Berre,
        and~Martin~Monperrus,~\IEEEmembership{Member,~IEEE,} 
\IEEEcompsocitemizethanks{
  \IEEEcompsocthanksitem J. Xuan is with the State Key Lab of Software Engineering, School of Computer, Wuhan University, Wuhan, China. \protect E-mail: \ \ \ \ \ \ \ \ \ \ \ \ \ \ \ \ \ \ \ \ \ \ \ \ \ \ \ \ \ 
  jxuan@whu.edu.cn.
  \IEEEcompsocthanksitem M. Martinez is with the University of Lugano, Lugano, Switzerland. He was with the University of Lille \& INRIA, Lille, France, when this work is done. \protect E-mail: matias.sebastian.martinez@usi.ch. 
  \IEEEcompsocthanksitem F. DeMarco and S. Lamelas Marcote are with the University of Buenos Aires, Buenos Aires, Argentina. \protect E-mail: faviod@gmail.com, srlm@gmx.com. 
  \IEEEcompsocthanksitem M. Cl\'{e}ment and T. Durieux are with the University of Lille, Lille, France. \protect E-mail: maxime.clement@etudiant.univ-lille1.fr, thomas.durieux@inria.fr.  
  \IEEEcompsocthanksitem D. Le Berre is with the University of Artois \& CNRS, Lens, France. \protect E-mail: leberre@cril.fr.
  \IEEEcompsocthanksitem M. Monperrus is with the University of Lille \& INRIA, Lille, France. \protect E-mail: martin.monperrus@univ-lille1.fr.
}

\thanks{$^{\dag}$\ F. DeMarco, M. Cl\'{e}ment, S. Lamelas Marcote, and T. Durieux have contributed to this work during their internship at INRIA Lille -- Nord Europe.}
\thanks{Manuscript received xxx; revised yyy}
}

\markboth{IEEE Transactions on Software Engineering}{Nopol}

\IEEEcompsoctitleabstractindextext{
\begin{abstract}

We propose \nopol, an approach to automatic repair of buggy conditional statements (i.e., \mycode{if-then-else} statements). This approach takes a buggy program as well as a test suite as input and generates a patch with a conditional expression as output. The test suite is required to contain passing test cases to model the expected behavior of the program and at least one failing test case that reveals the bug to be repaired. The process of \nopol consists of three major phases. First, \nopol employs angelic fix localization to identify expected values of a condition during the test execution. Second, runtime trace collection is used to collect variables and their actual values, including primitive data types and objected-oriented features (e.g., nullness checks), to serve as building blocks for patch generation. Third, \nopol encodes these collected data into an instance of a Satisfiability Modulo Theory (SMT) problem; then a feasible solution to the SMT instance is translated back into a code patch. 
We evaluate \nopol on \numbug real-world bugs (16 bugs with \buggyconditions and 6 bugs with \preconditions) on two large open-source projects, namely Apache Commons Math and Apache Commons Lang. Empirical analysis on these bugs shows that our approach can effectively fix bugs with \buggyconditions and \preconditions. We illustrate the capabilities and limitations  of \nopol using case studies of real bug fixes.

\end{abstract}
\begin{keywords}
Automatic repair, patch generation, SMT, fault localization
\end{keywords}
}

\maketitle

\IEEEdisplaynotcompsoctitleabstractindextext

\IEEEpeerreviewmaketitle

\section{Introduction}
\IEEEPARstart{A}{utomatic} software repair aims to automatically fix bugs in programs.
Different kinds of techniques are proposed for automatic repair, including patch generation \cite{le2012genprog, pei2014automated} and dynamic program state recovery \cite{Perkins2009, carzaniga2013automatic}.

A family of techniques has been developed around the idea of ``test-suite based repair'' \cite{le2012genprog}.
The goal of test-suite based repair is to generate a patch that makes failing test cases pass and keeps the other test cases satisfied. Recent test-suite based repair approaches include the work by Le~Goues et al. \cite{le2012genprog}, Nguyen et al. \cite{nguyen2013semfix}, Kim et al. \cite{Kim2013}. 

In recent work \cite{Martinez2013}, we have shown that \ourif conditions are among the most error-prone program elements in Java programs. 
In our dataset, we observed that 12.5\% of one-change commits simply update an \ourif condition. \emph{This motivates us to study the automatic repair of conditional statements in real-world bugs}. 

In this paper, we present a novel automatic repair system called \nopol.\footnote{\nopol is an abbreviation for ``no polillas'' in Spanish, which literally means ``no moth anymore''.} This system fixes conditional bugs in object-oriented programs and is evaluated on real bugs from large-scale open-source programs. For instance, \nopol can synthesize a patch that updates a buggy \ourif condition as shown in Fig. \ref{fig:example-buggy} or adds a guard precondition as in Fig. \ref{fig:example-pre}. Both figures are excerpts of real-world bugs taken from the bug tracking system of Apache Commons Math.
 
\nopol takes a buggy program as well as a test suite as input and generates a conditional patch as output. This test suite must contain at least one failing test case that embodies the bug to be repaired. Then, \nopol analyzes program statements that are executed by failing test cases to identify the source code locations where a patch may be needed.

For each statement, the process of generating the patch consists of three major phases. 
First, we detect whether there exists a fix location for a potential patch in this statement with a new and scalable technique called ``angelic fix localization'' (Section \ref{sect:angelic-fix-localization}). For one fix location, this technique reveals angelic values, which make all failing test cases pass.

\begin{figure}[!t]
\centering
\noindent\begin{minipage}{0.45\textwidth}
\begin{lstlisting}
-  if (u * v == 0) { 
+  if (u == 0 || v == 0) {
    return (Math.abs(u) + Math.abs(v));
  }  
\end{lstlisting}
\end{minipage}
\caption{Patch example of Bug CM5: a bug related to a \buggycondition. The original condition with a comparison with \mycode{==} is replaced by a disjunction between two comparisons.}
\label{fig:example-buggy}
\end{figure}

\begin{figure}[!t]
\centering
\noindent\begin{minipage}{0.45\textwidth}
\begin{lstlisting}
+ if (specific != null) {
    sb.append(": "); //sb is a string builder in Java
+ }
\end{lstlisting}
\end{minipage}
\caption{Patch example of Bug PM2: a precondition is added to avoid a null dereference.}
\label{fig:example-pre}
\end{figure}

Second, \nopol collects runtime traces from test suite execution through code instrumentation (Section \ref{sect:data-collection}). These traces contain snapshots of the program state at all candidate fix locations. The collected trace consists of both primitive data types (e.g., integers and booleans) and object-oriented data (e.g., nullness or object states obtained from method calls). 

Third, given the runtime traces, the problem of synthesizing a new conditional expression that matches the angelic values is translated into a Satisfiability Modulo Theory (SMT) problem (Section \ref{subsect:encoding}).
Our encoding extends the technique by Jha et al. \cite{jha2010oracle} by handling rich object-oriented data.
We use our own implementation of the encoding together with an off-the-shelf SMT solver (Z3 \cite{z3}) to check whether there exists a solution.

If such a solution exists, \nopol translates it back to source code, i.e., generates a patch. We re-run the whole test suite to validate whether this patch is able to make all test cases pass and indeed repairs the bug under consideration.

To evaluate and analyze our repair approach \nopol, we collect a dataset of \numbug bugs (16 bugs with \buggyconditions and 6 bugs with \preconditions) from real-world projects. Our result shows that 17 out of \numbug bugs can be fixed by \nopol, including four bugs with manually added test cases. Four case studies are conducted to present the benefits of generating patches via \nopol and five bugs are employed to explain the limitations.

The main contributions of this paper are as follows. 
\begin{itemize} 
\item The design of a repair approach for fixing conditional statement bugs of the form of \buggyandpres.
\item Two algorithms of angelic fix localization for identifying potential fix locations and expected values.
\item An extension of the SMT encoding in \cite{jha2010oracle} for handling nullness and certain method calls of object-oriented programs.
\item An evaluation on a dataset of \numbug bugs in real-world programs with an average of 25K executable lines of code for each bug.
\item A publicly-available system for supporting further replication and research. 
\item An analysis of the repair results with respect to fault localization. 
\end{itemize}

This paper is an extension of our previous work \cite{demarco2014automatic}. This extension adds an evaluation on a real-world bug dataset, four detailed case studies, a discussion of the limitations of our approach, and a detailed analysis on patches. 

The remainder of this paper is organized as follows. Section \ref{background} provides the background of test-suite based repair. Section \ref{sect:approach} presents our approach for repairing bugs with \buggyconditions and \preconditions. Section \ref{sect:evaluation} details the evaluation on \numbug real-world bugs. Section \ref{sect:discussions} further analyzes the repair results. Section \ref{sect:threats} presents potential issues and Section \ref{sect:relatedwork} lists the related work. Section \ref{sect:conclusion} concludes.

\section{Background}
\label{background}

We present the background on test-suite based repair and the two kinds of bugs targeted in this paper. 

\subsection{Test-Suite Based Repair}
\label{sect:test-suite-based-repair}

Test-suite based repair consists in repairing programs according to a test suite, which contains both passing test cases as a specification of the expected behavior of the program and at least one failing test case as a specification of the bug to be repaired.
Failing test cases can either identify a regression bug or reveal a new bug that has just been discovered.
Then, a repair algorithm searches for patches that make all the test cases pass.

The core assumption of test-suite based repair is that the test suite is good enough to thoroughly model the program to repair \cite{monperrus2014critical}. This is a case when the development process ensures a very strong programming discipline. For example, most commits of Apache projects (e.g., Apache Commons Lang) contain a test case specifying the change. If a commit is a bug fix, the commit contains a test case that highlights the bug and fails before the fix.

Test-suite based repair, which has been popularized by the work of GenProg by Le~Goues et al. \cite{le2012genprog}, has become an actively explored research area \cite{nguyen2013semfix,Kim2013,qi2014strength,qi2015efficient,DBLP:conf/sigsoft/LongR15}.
The approach presented in this paper, \nopol, is also an approach to test-suite based repair. Other kinds of repair methods include repair based on formal models \cite{Jobstmann2005} and dynamic repair of the program state \cite{Perkins2009}.

\subsection{Buggy \ourif Condition Bugs}

Conditional statements (e.g., \mycode{if (condition)\{\ldots\} else \{\ldots\}}), are widely-used in programming languages. Pan et al. \cite{pan2009toward} show that among seven studied Java projects, up to 18.6\% of bug fixes have changed a buggy condition in \ourif statements. A buggy \ourif condition is defined as a bug in the condition of an \mycode{if-then-else} statement. 

The bug in Fig. \ref{fig:example-buggy} is a real example of a buggy \ourif condition in Apache Commons Math. This bug is a code snippet of a method that calculates the greatest common divisor between two integers. The condition in that method is to check whether either of two parameters \mycode{u} and \mycode{v} is equal to 0. In the buggy version, the developer compares the product of the two integers to zero. However, this may lead to an arithmetic overflow. A safe way to proceed is to compare each parameter to zero. This bug was fixed by \nopol (see Bug CM5 in Table \ref{tab:patch}).

\subsection{Missing Precondition Bugs}
Another class of common bugs related conditions is the class of missing preconditions. 
A precondition aims to check the state of certain variables before the execution of a statement.
Examples of common preconditions include detecting a null pointer or an invalid index in an array.  
In software repositories, we can find commits that add preconditions (i.e., which were previously missing).

The bug in Fig. \ref{fig:example-pre} is a missing precondition with the absence of null pointer detection. The buggy version without the precondition throws an exception signaling a null pointer at runtime. \nopol fixed this bug by adding the precondition (see Bug PM2 in Table~\ref{tab:patch}).

\section{Our Approach}
\label{sect:approach}
This section presents our approach to automatically repairing buggy \ourif conditions and missing preconditions.
Our approach is implemented in a tool called \nopol that repairs Java code.

\subsection{Overview}
\label{subsect:overview}

\nopol is a repair approach, which is dedicated to buggy \ourif conditions and missing preconditions. As input, \nopol requires a test suite which represents the expected program functionality with at least one failing test case that exposes the bug to be fixed. Given a buggy program and its test suite, \nopol returns the final patch as output. Fig. \ref{fig:example-buggy} and Fig. \ref{fig:example-pre} are two examples of output patches for \buggyandpres by \nopol, respectively.

\textbf{How to use {\sc \textbf{Nopol}}}. From a user perspective, given a buggy program with a test suite, including failing test cases, the user would run \nopol and obtain a patch, if any. Before applying \nopol to the buggy program, the user does not need to know whether the bug relates to conditions. Instead, the user runs \nopol for any buggy program. 
If \nopol finds a patch, then the user would manually inspect and validate it before the integration in the code base. As further discussion in Section \ref{subsect:general-repair}, the user can also add a pre-defined timeout, e.g., 90 seconds as suggested in experiments or a longer timeout like five hours instead of exhaustively exploring the search space.

\begin{figure}[!t]
\centering
\includegraphics[width=0.5\textwidth]{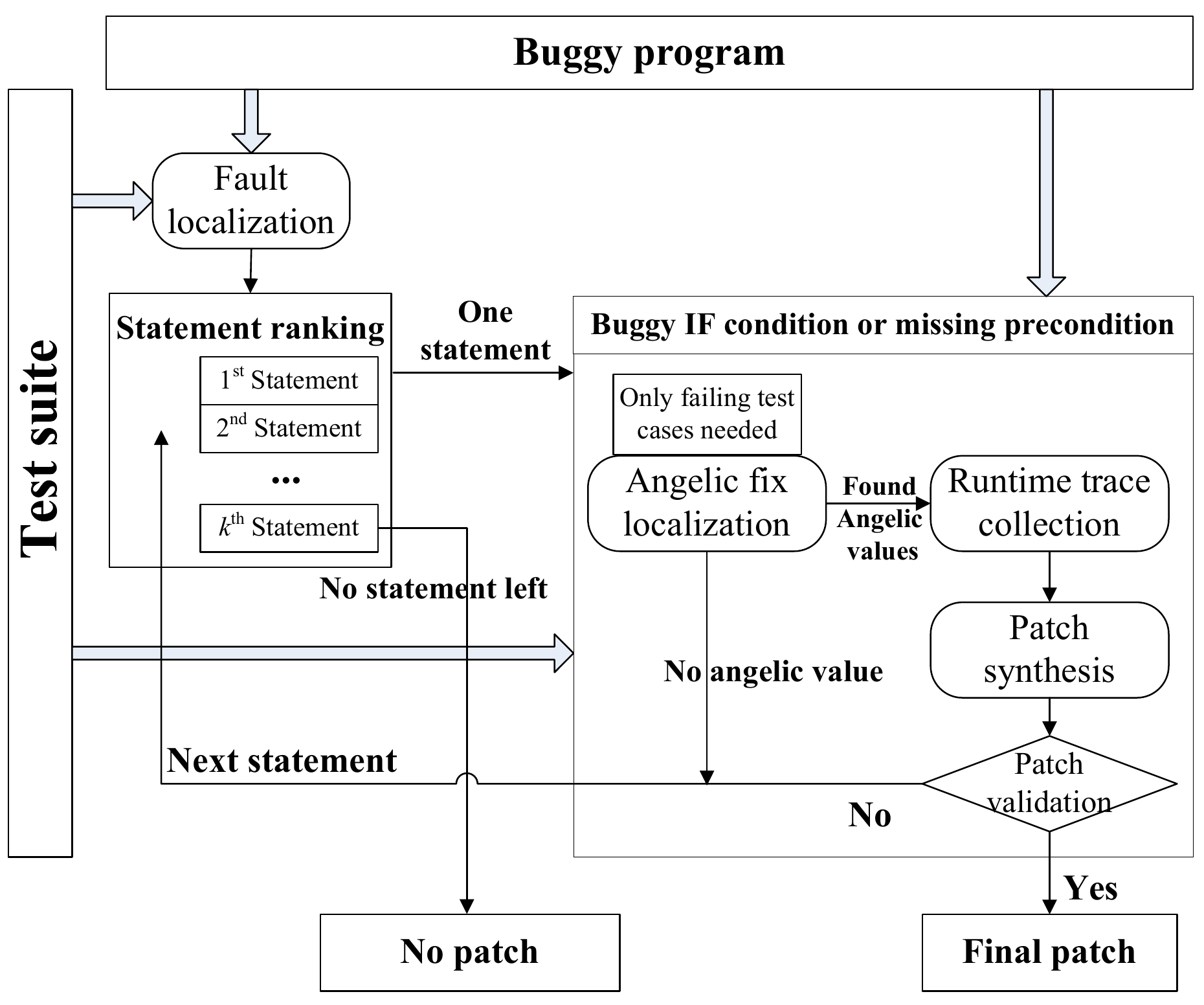}
\caption{Overview of the proposed automatic repair approach, \nopol.}
\label{fig:framework}
\end{figure}

Fig. \ref{fig:framework} shows the overview of \nopol.  
 \nopol employs a fault localization technique to rank statements according to their suspiciousness of containing bugs. For each statement in the ranking, \nopol considers it as a \buggycondition candidate if the statement is an \ourif statement; or \nopol considers it as a \precondition candidate if the statement is any other non-branch or non-loop statement (e.g., an assignment or a method call). \nopol processes candidate statements one by one with three major phases.

First, in the phase of \emph{angelic fix localization}, \nopol arbitrarily tunes a conditional value (\mycode{true} or \mycode{false}) of an \ourif statement to pass a failing test case. If such a conditional value exists, the statement is identified as a fix location and the arbitrary value is viewed as the expected behavior of the patch. In \nopol, there are two kinds of fix locations, \ourif statements for repairing buggy conditions and arbitrary statements for repairing missing preconditions.

Second, in the phase of \emph{runtime trace collection}, \nopol runs the whole test suite in order to collect the execution context of each fix location. The context includes both variables of primitive types (e.g., booleans or integers) and a subset of object-oriented features (nullness and certain method calls); then such runtime collection will be used in synthesizing the patch in the next phase. 

Third, in the phase of \emph{patch synthesis}, the collected trace is converted into a Satisfiability Modulo Theory (SMT) formula. The satisfiability of SMT implies that there exists a program expression that preserves the program behavior and fixes the considered bug. That is, the expression makes all test cases pass. If the SMT formula is satisfiable, the solution to the SMT is translated as a source code patch; if unsatisfiable, \nopol goes to the next statement in the statement ranking, until all statements are processed.

After the above three phases, the whole test suite is re-executed to validate that the patch is correct. This validation could be skipped if the SMT encoding is proven to be correct.
Indeed, theoretically, if the SMT solver says ``satisfiable'', it means that a patch exists. 
However, there could be an implementation bug in the trace collection, in the SMT problem generation, in the off-the-shelf SMT solver, or in the patch synthesis.  
Consequently, we do make the final validation by re-executing the test suite.

\subsection{Angelic Fix Localization}
\label{sect:angelic-fix-localization}

In \nopol, we propose to use value replacement \cite{jeffrey2008fault} to detect potential fix locations.
Value replacement \cite{jeffrey2008fault} comes from fault localization research. It consists in replacing at runtime one value by another one. 
More generally, the idea is to artificially change the program state for localizing faults. 
There are a couple of papers that explore this idea. 
For instance, Zhang et al. \cite{zhang2006locating} use the term ``predicate switching'' and Chandra et al. \cite{chandra2011angelic} use the term ``angelic debugging''. 

\nopol replaces conditional values in \ourif statements. We refer to conditional values that make test cases pass as \emph{angelic values}. 

{\bf Definition (Angelic Value)}
An angelic value is an arbitrarily-set value at a given location during test execution, which enables a failing test case to pass. 

To facilitate the description of our approach, we follow existing work \cite{gulwani2011synthesis} to introduce the concept of locations. A \textit{location} is an integer value, which identifies the absolute position of a statement in source code.

{\bf Definition (Angelic Tuple)}
An angelic tuple is a triplet $(loc, val, test)$, where the statement at a location $loc$ is evaluated to a value $val$ to make a failing test case $test$ pass.

In this paper, we refer to the technique of modifying the program state to find the values for angelic tuples $(loc, val, test)$ as \emph{angelic fix localization}. If an angelic tuple $(loc, val, test)$ is found, there may exist a patch in the location $loc$ in source code. In the phase of angelic fix localization, only failing test cases are needed, not the whole test suite. 

A single test case $test$ may evaluate the statement at the location $loc$ several times. Consequently, according to our definition, the value $val$ is fixed across all evaluations of a given statement for one test case. This is the key point for having a tractable search space (will be discussed in Section \ref{subsubsect:search-space}). On the other hand, one angelic value is specific to a test case: for a given location $loc$, different failing test cases may have different angelic values.

\subsubsection{For Buggy \ourif Conditions}

\begin{algorithm}[!t]

\SetKwInOut{Input}{Input}
\SetKwInOut{Output}{Output}

\Input{ \\
\noindent $stmt$, a candidate \ourif statement; \\
\noindent $T_f$, a set of failing test cases. 
}
\Output{ \\
\noindent $R$, a set of angelic tuples. 
}

\BlankLine 
$R \leftarrow \emptyset$\;
Initialize two sets $T_{true} \leftarrow \emptyset$ and $T_{false} \leftarrow \emptyset$\; 
Let $cond$ be the condition in $stmt$ and let $loc$ be the location of $cond$\; 

\BlankLine 
Force $cond$ to \mytrue and execute test cases in $T_f$\; 
\ForEach{failing test case $t_i \in T_f$}{
  \If{$t_i$ passes}{
    $T_{true} \leftarrow T_{true} \cup \{t_i\}$\;
  }
}
Force $cond$ to \myfalse and execute test cases in $T_f$\; 
\ForEach{failing test case $t_i \in T_f$}{
  \If{$t_i$ passes}{
    $T_{false} \leftarrow T_{false} \cup \{t_i\}$\;
  }
}

\BlankLine 
\tcp{All test cases in $T_f$ are passed}
\If{$(T_f \setminus T_{true}) \cap (T_f \setminus T_{false}) = \emptyset$}{
  \ForEach{$t_i \in T_{true}$}{
    $R \leftarrow R \cup \{(loc, \mbox{\mytrue}, t_i)\}$\;
  }
  \ForEach{$t_i \in T_{false}$}{
    $R \leftarrow R \cup \{(loc, \mbox{\myfalse}, t_i)\}$\;
  }
}
\BlankLine 
\caption{Angelic Fix Localization Algorithm for Buggy \ourif Conditions}
\label{alg:if-angelic-fix-localization}
\end{algorithm}

For buggy \ourif conditions, angelic fix localization works as follows.
For each \ourif condition that is evaluated during test suite execution, an angel forces the \ourif condition to be \mycode{true} or \mycode{false} in a failing test case. 
An angelic tuple $(loc, val, test)$, i.e., (\ourif condition location, boolean value, failing test case), indicates that a fix modifying this \ourif condition may exist (if the subsequent phase of patch synthesis succeeds, see Section \ref{subsect:encoding}).

Algorithm \ref{alg:if-angelic-fix-localization} is the pseudo-code of angelic fix localization for \buggyconditions. For a given \ourif statement \textit{stmt} and its condition \textit{cond}, both \mytrue and \myfalse are set to pass originally failing test cases at runtime. Lines 4 to 9 and Lines 10 to 15 describe how to set \textit{cond} to be \mytrue and \myfalse, respectively. If all failing test cases are passed, angelic tuples are collected, i.e., Lines 17 to 22, for further patch synthesis; otherwise, 
there exists no angelic value for the test case and the location under consideration.  
The same idea of forcing the execution can be used to identify angelic values for loop conditions \cite{Lamelas2015}.

\subsubsection{For Missing Preconditions}
\label{subsubsect:precondition}

Angelic fix localization for \preconditions is slightly different from that for \ourif conditions.
For each non-branch and non-loop statement that is evaluated during test suite execution, an angel forces to skip it. 
If a failing test case now passes, it means that a potential fix location has been found. 
The oracle for repair is then  ``false''; that is, the added precondition must be \myfalse, i.e., the statement should be skipped. Then, an angelic tuple $(loc, val, test)$ is (precondition location, \mycode{false}, failing test case). 

Algorithm \ref{alg:missing-pred-angelic-fix-localization} is the pseudo-code of this algorithm. Given a non-\ourif statement \textit{stmt}, we skip this statement to check whether all failing test cases are passed, i.e., Lines 4 to 9. If yes, the location \textit{loc} of the precondition as well as its angelic value \mycode{false} is collected, i.e., Lines 11 to 13. 
If skipping the statement does not pass all the failing test cases, no angelic values will be returned. 
This technique also works for \preconditions for entire blocks since blocks are just specific statements in Java. In our implementation, we only consider adding missing preconditions for single statements rather than blocks. Manual examination on the dataset in Section \ref{subsect:general-repair} will show that our dataset does not contain missing preconditions for blocks. 

\begin{algorithm}[!t]
\SetKwInOut{Input}{Input}
\SetKwInOut{Output}{Output}

\Input{ \\
\noindent $stmt$, a candidate non-\ourif statement; \\
\noindent $T_f$, a set of failing test cases. 
}
\Output{ \\
\noindent $R$, a set of angelic tuples. 
}

\BlankLine  
$R \leftarrow \emptyset$\;
Initialize a test case set $T_{pre} \leftarrow \emptyset$\; 
Let $loc$ be the location of a potential precondition of $stmt$;

\BlankLine
Force $stmt$ to be skipped and execute $T_f$\; 
\ForEach{failing test case $t_i \in T_f$}{
  \If{$t_i$ passes}{
    $T_{pre} \leftarrow T_{pre} \cup \{t_i\}$\;
  }
}

\BlankLine
\tcp{All test cases in $T_f$ are passed}
\If{$T_{pre} = T_f$}{
  \ForEach{$t_i \in T_{pre}$}{
    $R \leftarrow R \cup \{(loc, \mbox{\myfalse}, t_i)\}$\;
  }
}
\BlankLine
\caption{Angelic Fix Localization Algorithm for Missing Preconditions}
\label{alg:missing-pred-angelic-fix-localization}
\end{algorithm}

\subsubsection{Characterization of the Search Space}
\label{subsubsect:search-space}
We now characterize the search space of angelic values. 
If an \ourif condition is executed more than once in a failing test case, there may exist a sequence of multiple different angelic values resulting in a passing test case. For example, a \buggycondition that is executed three times by one failing test case may require a sequence of three different angelic values to pass the test case. 

\textbf{Search space for \buggyconditions}. 
In general, if one failing test case executes a \buggycondition for $t_c$ times, the search space of all sequences of angelic values is $2^{t_c}$.
To avoid the problem of combinatorial explosion, \nopol assumes that, for a given failing test case, the angelic value is the same during the multiple executions on one statement. 
The search space size becomes $2$ for one failing test case instead of $2^{t_c}$. Under this assumption, the search space is shown as follows. 

For \buggycondition, the search space is $2\times n_c$ where $n_c$ is the number of executed \ourif statements by a given failing test case.

\textbf{Search space for \preconditions}.
Similarly to angelic fix localization for buggy \ourif conditions, if a statement is executed several times by the same failing test case, angelic fix localization directly adds a precondition (with a \mycode{false} value) and completely skips the statement for a given test case. 

For \precondition bugs, the search space size is $n_p$, where $n_p$ is the number of executed statements by test cases. It is not $2\times n_p$ because we only add a precondition and check whether the \mycode{false} value passes the failing test case.

\nopol does not decide a priority between updating existing conditions or adding new preconditions. A user can try either strategy, or both. There is no analytical reason to prefer one or the other; our evaluation does not give a definitive answer to this question. In our experiment, we perform both strategies for statements one by one (see Section \ref{subsect:overview}).

If no angelic tuple is found for a given location, there are two potential reasons. 
First, it is impossible to fix the bug by changing the particular condition (resp. adding a precondition before the statement).
Second, only a sequence of different angelic values, rather than a single angelic value, would enable the failing test case to pass.
Hence, \nopol is incomplete: there might be a way to fix an \ourif condition by alternating the way of finding angelic values, but we have not considered it in this paper.

\subsection{Runtime Trace Collection for Repair}
\label{sect:data-collection} 

Once an angelic tuple is found, \nopol collects the values that are accessible at this location during program execution. Those values are used to synthesize a correct patch (in Section \ref{subsect:encoding}). In our work, different kinds of data are collected to generate a patch.

\subsubsection{Expected Outcome Data Collection}
\label{subsubsect:outcome}

As mentioned in Section \ref{sect:angelic-fix-localization}, an angelic value indicates that this value enables a failing test case to pass. To generate a patch, \nopol collects the expected outcomes of conditional values to pass the whole test suite: angelic values for failing test cases as well as actual execution values for the passing ones. 

Let $O$ be a set of expected outcomes in order to pass all test cases. An expected outcome $O_{loc,m,n} \in O$ refers to the value at location $loc$ during the $m$-th execution in order to pass the $n$-th test case. \nopol collects $O_{loc,m,n}$ for all executions of location $loc$.

\textbf{For buggy \ourif conditions}. $O_{loc,m,n}$ is the expected outcome of the condition expression at $loc$.
For a failing test case, the expected outcome is the angelic value; 
for a passing test case, the expected outcome is the runtime value $eval(loc)$, i.e., the result of the evaluation during the actual execution of the \ourif condition expression.
$$ O_{loc,m,n} =
\begin{cases}
eval(loc), & \mbox{for passing test cases} \\
\mbox{angelic value} & \mbox{for failing test cases} 
\end{cases}
$$

\textbf{For missing preconditions}. $O_{loc,m,n}$ is the expected value of the precondition at $loc$, i.e., \mycode{true} for passing test cases and \mycode{false} for failing test cases. The latter comes from angelic fix localization: if the precondition returns \mycode{false} for a failing test case, the buggy statement is skipped and the test case passes.
$$ O_{loc,m,n} =
\begin{cases}
true & \mbox{for passing test cases} \\
false & \mbox{for failing test cases} 
\end{cases}
$$
  
Note that not all the bugs with missing preconditions can be fixed with the above definition. Section \ref{subsubsect:limit-pl3} will present the limitation of this definition with a real example.    

\subsubsection{Primitive Type Data Collection}
\label{subsect:primitive}

At the location of an angelic tuple, \nopol collects the values of all local variables, method parameters, and class fields that are typed with a basic primitive type (booleans, integers, floats, and doubles).

Let $C_{loc,m,n}$ be the set of collected values at location $loc$ during the $m$-th execution of the $n$-th test case.
In order to synthesize conditions that use literals (e.g., \mycode{if (x > 0)}), $C_{loc,m,n}$ is enriched with constants for further patch synthesis.
First, \nopol collects static values that are present in the program.\footnote{Besides collecting static fields in a class, we have also tried to collect other class fields of the class under repair, but the resulting patches are worse in readability than those without collecting class fields. Hence, no class fields other than static fields are involved in data collection. } Second, we add three standard values  \{0, -1, 1\}, which are present in many bug fixes in  the wild (for instance for well-known off-by-one errors).
Based on these standard values, other values can be formed via wiring building blocks (in Section \ref{subsubsect:wiring}). For example, a value $2$ in a patch can be formed as $1+1$, if $2$ is not collected during runtime trace collection. 

\subsubsection{Object-Oriented Data Collection}
\label{subsubsect:oo}

\nopol aims to support automatic repair for object-oriented programs. 
In particular, we would like to support nullness checks and some particular method calls.
For instance, \nopol is able to synthesize the following patch containing a method call. 

\begin{lstlisting}
  +  if (obj.size() > 0) {
       compute(obj);
  +  }
\end{lstlisting}

To this end, in addition to collecting all values of primitive types, \nopol collects two kinds of object-oriented features.
First, \nopol collects the nullness of all variables of the current scope.
Second, \nopol collects the output of ``state query methods'', defined as the methods that inspect the state of objects and are side-effect free. A state query method is an argumentless method with a primitive return type. 
For instance, methods \mycode{size()} and \mycode{isEmpty()} of \mycode{Collection} are state query methods.
The concept of state query methods is derived from ``argument-less boolean queries'' of  ``object states'' by Pei et al. \cite{pei2014automated}.  

\nopol is manually fed with a list of such methods. The list is set with domain-specific knowledge. For instance, in Java, it is easy for developers to identify such side-effect free state query methods on core library classes such as \mycode{String}, \mycode{File} and \mycode{Collection}. For each object-oriented class $T$, those predefined state query methods are denoted as $sqm(T)$.

\nopol collects the nullness of all visible variables and the evaluation of state query methods for all objects in the scope (local variables, method parameters, and fields) of a location where an angelic tuple exists. Note that this incorporates inheritance; the data are collected based on the polymorphism in Java.
For instance, when the value of \texttt{obj.size()} is collected, it may be for one implementation of \texttt{size()} based on array lists and for another implementation of \texttt{size()} based on linked lists. This means that a patch synthesized by \nopol can contain polymorphic calls. 

\subsubsection{On the Size of Collected Data}

Let us assume there are $u$ primitives values and a set $O$ of $w$ objects in the scope of an angelic tuple. 
In total, \nopol collects the following values:

\begin{itemize}
\item $u$ primitive variables in the scope;
\item $w$ boolean values corresponding to the nullness of each object;
\item $\sum_{o \in O}|sqm(class(o))|$ values corresponding to the evaluation of the state query methods of all objects available in the scope, where $class(o)$ denotes the class of the object $o$;
\item constants, i.e., 0, -1, and 1 in our work.
\end{itemize}

\subsection{Patch Synthesis: Encoding Repair in SMT}
\label{subsect:encoding}

The patch synthesis of buggy \ourif conditions and missing preconditions consists of synthesizing an expression $exp$ such that
\begin{equation}
\forall_{loc,m,n} \mbox{~~}exp(C_{loc,m,n}) =  O_{loc,m,n} 
\label{eq:repair-if}
\end{equation}

As a synthesis technique, \nopol, as SemFix \cite{nguyen2013semfix}, uses a variation of oracle-guided component-based program synthesis \cite{jha2010oracle}, which is based on SMT.

The solution of the SMT problem is then translated back to a boolean source code expression $exp$ representing the corrected \ourif condition or the added precondition.

\subsubsection{Building Blocks}
\label{sect:bb}

We define a {\em building block} (called a {\em component} in \cite{jha2010oracle}) as a type of expression that can appear in the boolean expression to be synthesized. 
For instance, the logical comparison operator ``<'' is a building block. As building block types, we consider comparison operators ($<$, $>$, $\leq$, $\geq$, $=$, and $\neq$), arithmetic operators ($+$, $-$, and $\times$),\footnote{Adding the division is possible but would require specific care to avoid division by zero.} and boolean operators ($\wedge$, $\vee$, and $\lnot$). The same type of building blocks can appear multiple times in one expression. 

Let $b_i$ be the $i$th building block ($i=1,2,\ldots,k$). Then $b_i$ is identified as a tuple of a set of input variables $I_i$, an output variable $r_i$, and an expression $\phi_{i}(I_i,r_i)$ encoding the operation of the building block.
That is $b_i=(\phi_{i}(I_i,r_i),I_i,r_i)$ ($b_i=(\phi_{i},I_i,r_i)$ for short). For example, given a boolean output $r_i$ and an input $I_i=\{I_{i,1}, I_{i,2}\}$ consisting of two boolean values, a building block could be $b_i=(\phi_{i},\{I_{i,1}, I_{i,2}\},r_i)$, where $\phi_{i}$ is implemented with the operator $\wedge$, i.e.,  $I_{i,1} \wedge I_{i,2}$. 

Let $r$ be the final value of the synthesized patch. Hence there exists one building block $b_i$ whose output is bound to the return value $r_i = r$.

Suppose we are given a set $B$ of building blocks and a list $CO$ of pairs $(C_{loc,m,n},O_{loc,m,n})$, i.e., pairs of collected values and expected values at the location $loc$ during the $m$-th execution of the $n$-th test case. $C_{loc,m,n}$ includes values of different types: BOOL, INT, or REAL.\footnote{In the context of SMT, we use BOOL, INT, and REAL to denote the types of booleans, integers, and doubles as well as floats in Java, respectively.} A patch is a sequence of building blocks $<b_1,b_2,...,b_k>$ with $b_i \in B$, whose input values are taken from either $C_{loc,m,n}$ or other building blocks. 

\subsubsection{Wiring Building Blocks}
\label{subsubsect:wiring}

The problem of patch synthesis is thus to wire the input of building blocks $<b_1,b_2,...,b_k>$ to the input values $I_0$ or the output values of other building blocks. To synthesize a condition, we need to make sure that the types of the variables are valid operands (e.g., an arithmetic operator only manipulates integers). 

\textbf{Example}. Let us assume that $C_{loc,m,n}$ has three values, an integer variable $i_0$, a boolean constant $c_1 \leftarrow False$, and an integer constant $c_2 \leftarrow 3$. Assume we have two building blocks,  $\mbox{BOOL} \leftarrow f_1(\mbox{BOOL})$ and 
$\mbox{BOOL} \leftarrow f_2(\mbox{INT}, \mbox{INT})$. 
Then the goal of patch synthesis is to find a well formed expression, such as $False$, $f_1(False)$, $f_2(i_0, 3)$, $f_2(3,3)$, and $f_1(f_2(3,3))$; meanwhile, one of these expressions is expected to match the final output $r$.

\subsubsection{Mapping Inputs and Outputs with Location Variables}

Let $I = \cup I_i$ and $O = \cup \{r_i\}$ be the sets of input and output values of all building blocks $b_i \in B$. 
Let $I_0$ be the input set $\{ C_{loc,m,n} \}$ and let $r$ be the output in the final patch.
We define $IO$ as $IO = I \cup O \cup I_0 \cup \{ r \}$. 
We partition the elements of $IO$ according to their types in BOOL, INT, and REAL.

The SMT encoding relies on the creation of location variables. A {\em location variable} $l_x \in L$ represents an index of an element $x\in IO$. Note that the concept of location variables in SMT encoding is different from the concept of locations in Section \ref{sect:angelic-fix-localization}. A location variable $l_x$ indicates a relative position, i.e., an index, of $x$ in a patch while a location $loc$ indicates an absolute position of a statement in a source file. 
A {\em value variable} $v_x \in V$ represents a value taken by an elements $x \in IO$. Values of location variables are actually integers ($L \subseteq \mbox{INT}$);
value variables are of any supported type, i.e., BOOL, INT, or REAL. 

Informally, a location variable $l_x$ serves as an index of $x \in IO$ in a code fraction while a value variable $v_x$ indicates its value during test execution. Section \ref{subsubsect:domain} will further illustrate how to index the code via location variables. 
Location variables are invariants for the execution of all test cases: they represent the patch structure.
Value variables are used internally by the SMT solver to ensure that the semantics of the program is preserved.

\subsubsection{Domain Constraints}
\label{subsubsect:domain}

Let us first define the domain constraints over the location variables. Given the input set $I_0$ and the building block set $B$, let $p$ be the number of possible inputs and $p=|I_0|+|B|$. 
The location variables of the elements of $I_0$ and $r$ are fixed:
$$\phi_{FIXED}(I_0,r) = ( \land_{i=1}^{|I_0|} \ l_{I_0,i} = i ) \bigwedge l_r = p$$

Given building blocks $b_i \in B$, the location variable $l_{r_i}$ for the output $r_i$ ($r_i \in O$) of $b_i$ belongs to a range of $[|I_0|+1, p]$:
$$\phi_{OUTPUT}(O) = \bigwedge_{i=1}^{|O|} ( |I_0|+1 \leq l_{r_i} \leq p)$$

\textbf{Handling types}. Only the location variables corresponding to the values of the same type are allowed.
Suppose that $type(x)$ returns the set of elements with the same type of $x$ among BOOL, INT, and REAL. Then we can restrict the values taken by the location variables of the input values of building blocks using the following formula:

$$\phi_{INPUT}(I) = \bigwedge_{x \in I} \ \ \bigvee_{y \in type(x), x \neq y} (l_x = l_y)$$

Recall the example in Section \ref{subsubsect:wiring}, we have the input $I_0=\{i_0, c_1, c_2\}$, the output $r$, and two building blocks $f_1$ and $f_2$. We assume that each building block is involved in the patch for at most once for simplifying the example; in our implementation, a building block can be used once or more in a synthesized patch. Then we have the location variables as follows. The location variables of $i_0$, $c_1$, and $c_2$ are 1, 2, and 3; the locations variables of building blocks are 4 and 5, respectively. Based on the types, candidate values of location variables of $I_{f_1, 1}$, $I_{f_2, 1}$, and $I_{f_2, 2}$ are calculated. 

\begin{table}[H]
\centering
\resizebox{0.5\textwidth}{!}{
\begin{tabular}{l l}
$l_{i_0}=1$  &   input variable, integer  \\     
$l_{c_1} = 2 $ &  boolean constant, $False$ \\
$l_{c_2}  = 3$ &  integer constant, 3  \\
$l_{r_{f_1}} \in [4,5]$  &  output of $f_1$,  boolean  \\
$l_{r_{f_2}} \in [4,5]$  &  output of $f_2$, boolean \\ 
$l_r     = 5$    &  expected output value, boolean  \\
$l_{I_{f_1,1}} \in \{l_{c_1}, l_{r_{f_1}}, l_{r_{f_2}}\}$    &  the parameter of $f_1$, boolean \\
$l_{I_{f_2,1}}\in \{l_{i_0}, l_{c_2}\}$    &  the first parameter of $f_2$, integer \\
$l_{I_{f_2,2}} \in \{l_{i_0}, l_{c_2}\}$   &  the second parameter of $f_2$, integer \\
\end{tabular}
}
\end{table}

The following additional constraints are used to control the values of location variables.
First, we ensure that each output of a building block is mapped to one distinct input (wires are one-to-one).
$$\phi_{CONS}(L,O) = \bigwedge_{x,y \in O, x \neq y} l_x \neq l_y$$

Second, we need to order the building blocks in such a way that its arguments have already been defined.
$$\phi_{ACYC}(B,L,I,O) = \bigwedge_{(\phi_{i},I_i,r_i) \in B} \ \  \bigwedge_{x \in I_i} \ \  l_x < l_{r_i}$$
 
Then, we combine all constraints together. 
\begin{align*}
\phi_{WFF}(B,L,I,O, I_0,r) = \phi_{FIXED}(I_0,r)  \land \phi_{OUTPUT}(O) \\ \land \phi_{INPUT}(I) \land \phi_{CONS}(L,O) \land \phi_{ACYC}(B,L,I,O)
\end{align*}

An assignment of $L$ variables respecting the predicate $\phi_{WFF}(B,L,I,O,I_0,r)$ corresponds to a syntactically correct patch. 

Value variables corresponding to the input and the output of a building block are related according to the functional definition of a predicate $pb_i$. Given a building block $b_i = \{ \phi_i, I_i, r_i\}$, let $value(I_i)$ be a function that returns the value for the input $I_i$. For a value variable $v_{r_i}$, let $pb_i(value(I_i),v_{r_i})=true$ iff $\phi_i(I_i)=r_i$. Given $V_{IO} = \{v_x | x \in I \cup O\}$, we define the following constraint.  
$$\phi_{LIB}(B,V_{IO}) = \bigwedge_{(\phi_i, I_i, r_i) \in B, v_{r_i}\in V_{IO}}\ pb_i \Big(value(I_i),v_{r_i}\Big)$$

The location variables and the value variables are connected together using the following rule which states that elements at the same position should have the same value. Note that we need to limit the application of that rule to values of the same type because in our case, input or output values can be of different types. Such a limitation to the elements of the same type is valid since the domain of the location variables are managed using constraints $ \phi_{INPUT}(I)$. 
\begin{align*}
\phi_{CONN} & (L,V_{IO}) =     \\
  & \bigwedge_{S \in \{\mbox{\footnotesize BOOL},\mbox{\footnotesize INT},\mbox{\footnotesize REAL}\}}\ \bigwedge_{x,y \in S}\ l_x = l_y \Rightarrow v_x = v_y
\end{align*}

Let the notation $\alpha[v \leftarrow x]$ mean that the variable $v$ in the constraint $\alpha$ has been set to the value $x$. For a given location $loc$, the patch for a given input $I_0$ and a given output $r$ is preserved using the following existentially quantified constraint.
\begin{align*}
\phi&_{FUNC} (B,L,C_{loc,m,n},O_{loc,m,n}) = \\
  & \exists V_{IO} \Big( \phi_{LIB}(B,V_{IO}) \bigwedge  \\
  & \phi_{CONN}(L,V_{IO}) [value(I_0) \leftarrow  C_{loc,m,n}, v_r \leftarrow O_{loc,m,n}] \Big)
\end{align*}

Finally, finding a patch which satisfies all expected input-output pairs $(C_{loc,m,n},O_{loc,m,n})$ requires to satisfy the following constraint.
\begin{align*}
\phi&_{PATCH}(B,I,O,CO,I_0,r) =  \\
  \exists & L \Big( \bigwedge_{(C_{loc,m,n},O_{loc,m,n}) \in CO} \  \phi_{FUNC}(B,L,C_{loc,m,n},O_{loc,m,n}) \\
  & \bigwedge \ \phi_{WFF}(B,L,I,O,I_0,r)   \Big)
\end{align*}

\subsubsection{Complexity Levels of Synthesized Expressions}
\label{sec:smt-level}

Ideally, we could feed SMT with many instances of all kinds of building blocks (see Section \ref{sect:bb}). 
Only the required building blocks would be wired to the final result.
This is an inefficient strategy in practice: some building blocks require expensive computations, e.g., a building block for multiplication (which is a hard problem in SMT).

To overcome this issue, we use the same technique of complexity levels as SemFix \cite{nguyen2013semfix}.
We first try to synthesize an expression with only one instance of easy building blocks ($<$, $\leq$, $\neq$, and $=$)\footnote{$>$ and $\geq$ are obtained by symmetry, e.g., $a~\geq~b$ as $b~\leq~a$.}.
Then, we add new building blocks (e.g., building blocks of logical operators and arithmetic operators, successively) and eventually we increase the number of instances of building blocks. We refer to those successive SMT satisfaction trials as the ``SMT level''.

\subsubsection{Patch Pretty-Printing}
\label{subsubsect:patch-print}

\nopol translates a solution to a patch in source code if there is a feasible solution to the SMT problem. Since \nopol repairs bugs with \buggyandpres, the patch after translation is a conditional expression, which returns a boolean value. 

The translation is obtained with a backward traversal starting at the final output location $l_r$, as explained in Algorithm \ref{alg:patch-pretty-print}. A function $traverse$ returns the traversal result according to the location variables while a function $code$ converts a variable into source code. For example, for a variable $a$, $code(a)$ is translated to $a$; if $\phi$ denotes the conjunction of boolean values, $code(\phi(traverse(a),traverse(b)))$ is translated to $a \wedge b$. As shown in Algorithm \ref{alg:patch-pretty-print}, patch translation from a SMT solution to a patch is a deterministic algorithm, which generates an identical patch for a given SMT solution. Once a patch is translated from the SMT solution, \nopol returns this patch to developers as the final patch.   

\begin{algorithm}[!t]
\SetKwFunction{traverse}{traverse}

\SetKwInOut{Input}{Input}
\SetKwInOut{Output}{Output}

\Input{ \\
\noindent $L$, an assignment of location variables, i.e., an SMT solution; \\
\noindent $r$, a final and expected output variable of patch; \\
}
\Output{ \\
\noindent $patch$, a source code patch. 
}

\BlankLine  
Find a location variable $l_x = l_r$\;
$patch=$ \traverse{$l_x$}\; 
\BlankLine
\SetKwProg{myproc}{Function}{}{end}
\myproc{\traverse{$l_x$}}{
  \uIf(\ \ // Output of a building block){$x \in O$} 
  {
    Find the expression $\phi_x(I_x, x)$ and $I_x=(I_{x,1},I_{x,2},...)$\; 
    \KwRet \footnotesize $code(\phi_x(traverse(I_{x,1}),traverse(I_{x,2}),...))$\;  
  }
  \uElseIf(// Input of a building block){$x \in I$}
  {
    Find $y$ for $l_y = l_x$; \ \ // $l_y \in O \cup I_0$\\
    \KwRet $traverse(l_y)$\;
  }
  \Else(\ \ // $x \in I_0$, from collected runtime trace)
  {
    \KwRet $code(x)$\;
  }
}
\BlankLine
\caption{Translation Algorithm from an SMT Solution to a Source Code Patch.}
\label{alg:patch-pretty-print}
\end{algorithm}

Here is a possible solution to the SMT instance for our running example (in Section \ref{subsubsect:wiring}): 
\noindent 
$l_{i_0}   = 1$,              
$l_{c_1}   = 2    $,    
$l_{c_2}   = 3        $,
$l_{r}    = 5       $,
$l_{r_{f_1}}  = 4       $, 
$l_{r_{f_2}}  = 5       $, 
$l_{I_{f_1,1}}= 2  $,
$l_{I_{f_2,1}} = 1    $,
$l_{I_{f_2,2}} = 1  $.

In our example, the output is bound to $l_r=5$ that is the output of $f_2$. Then
$f_2$ takes the integer input value $i_0$ in $l_{i_0}$ as a parameter.
The final patch is thus the expression $f_2(i_0, i_0)$ which returns a boolean. This patch could be the repair of a bug, i.e., a fixed \ourif condition or an added precondition. In this example, $f_1$ is never used.

\subsection{Fault Localization}
\label{subsect:fault-localization}

\nopol uses an existing fault localization technique to speed up finding an angelic value, if one exists. In fault localization, statements are ranked according to their suspiciousness. The \textit{suspiciousness} of a statement measures its likelihood of containing a fault.

In \nopol, a spectrum-based ranking metric, Ochiai \cite{abreu2007accuracy}, is used as the fault localization technique. Existing empirical studies \cite{steimann2013threats,xuan2014learning} show that Ochiai is more effective on localizing the root cause of faults in object-oriented programs than other fault localization techniques. In Section \ref{subsect:discussion-fault}, we will compare the effectiveness among different fault localization techniques.  

Given a program and a test suite, the suspiciousness $susp(s)$ of a statement $s$ is defined as follows. 
\begin{equation*}
susp(s) =  \frac{failed(s)}{\sqrt{total\_failed*(failed(s)+passed(s))}}
\end{equation*}
where $total\_failed$ denotes the number of all the failing test cases and $failed(s)$ and $passed(s)$ respectively denote the number of failing test cases and the number of passing test cases, which cover the statement $s$. Note that $0 \leq susp(s) \leq 1$ where $susp(s)=1$ indicates the highest probability of localizing the bug and $susp(s)=0$ indicates there is no likelihood between this statement and the bug.  

We rank all the statements based on their suspiciousness in descending order. For all the statements with the suspiciousness over zero, we detect whether this statement is an \ourif statement or not. As previously mentioned in Section \ref{subsect:overview}, for an \ourif condition, \nopol tries to synthesize a new condition while for a non-\ourif statement, \nopol tries to add a precondition.

\section{Automatic Repair of Real-World If Bugs}
\label{sect:evaluation}

We now evaluate our repair approach, \nopol, on a dataset of \numbug real-world bugs. First, we describe our evaluation methodology in Section \ref{subsect:methodology}; second, we introduce the setup of our dataset in Section \ref{subsect:data-set} and the implementation details in Section \ref{subsect:impl}; third, we present the general description of the synthesized patches in Section \ref{subsect:general-repair}; fourth, four bugs are employed as case studies in Section \ref{subsect:case-study-repair} and five bugs are used to illustrate the limitations in Section \ref{subsect:limitation}.

\subsection{Evaluation Methodology}
\label{subsect:methodology}

Our evaluation methodology is based on the following principles. 

\textbf{P1}. We evaluate our tool, \nopol, on real-world buggy programs (Section \ref{subsect:general-repair}). 

\textbf{P2}. For bugs that \nopol can fix, we examine the automatically generated patches, and compare them with human-produced patches (Section \ref{subsect:case-study-repair}). 

\textbf{P3}. For bugs that \nopol cannot correctly fix, we check the details of these bugs and highlight the reasons behind the unrepairability (Section \ref{subsect:limitation}). When the root cause is an incorrect test case (i.e., an incomplete specification), we modify the test case and re-run \nopol.

\textbf{P4}. We deliberately do not compute a percentage of repaired bugs because this is a potentially unsound measure. According to our previous investigation \cite{monperrus2014critical}, this measure is sound if and only if 1) the dataset is only composed of bugs of the same kind and 2) the distribution of complexity within the dataset reflects the distribution of all in-the-field bugs within this defect class. In our opinion, the second point is impossible to achieve.

We have not quantitatively compared our approach against existing repair approaches on the same dataset because
1) either existing approaches are inapplicable on this dataset (e.g., GenProg \cite{le2012genprog} and SemFix \cite{nguyen2013semfix} are designed for C programs);
2) or these approaches are not publicly available (e.g., PAR \cite{Kim2013} and mutation-based repair \cite{debroy2010using}).

\subsection{Dataset of Real-World Bugs}
\label{subsect:data-set}

\nopol focuses on repairing conditional bugs, i.e., bugs in \ourif conditions and preconditions. Hence, we build a dataset of 22 real-world bugs of buggy \ourif conditions and missing preconditions. Since our prototype implementation of \nopol repairs Java code, these \numbug bugs are selected from two open-source Java projects, Apache Commons Math\footnote{Apache Commons Math, \url{http://commons.apache.org/math/}.} and Apache Commons Lang\footnote{Apache Commons Lang, \url{http://commons.apache.org/lang/}.} (\textit{Math} and \textit{Lang} for short, respectively).  

Both Math and Lang manage source code using Apache Subversion\footnote{Apache Subversion, \url{http://subversion.apache.org/}.} (SVN for short) and manage bug reports using Jira.\footnote{Jira for Apache, \url{http://issues.apache.org/jira/}.} Jira stores the links between bugs and related source code commits. In addition, these projects use the FishEye browser to inspect source code and commits.\footnote{FishEye for Apache, \url{http://fisheye6.atlassian.com/}.} 

In our work, we employ the following four steps to collect bugs for the evaluation. 
First, we automatically extract small commits that modify or add an \ourif condition using Abstract Syntax Tree (AST) analysis \cite{Falleri2014}.
We define a \textit{small commit} as a commit that modifies at most 5 files, each of which introduces at most 10 AST changes (as computed by the analytical method, GumTree \cite{Falleri2014}).
In Math, this step results in 161 commits that update \ourif conditions and 104 commits that add preconditions; in Lang, the commits are 165 and 91, respectively. The lists of commits are available at the \nopol project \cite{nopol2014}.
Second, for each extracted commit, we collect its related code revision, i.e., the source program corresponding to this commit. 
We manually check changes between the code revision and its previous one; we only accept changes that contain an \ourif condition or a missing precondition and do not affect other statements. Those commits could also contain other changes that relate to neither a bug nor a patch, such as a variable renaming or the addition of a logging statement. 
In this case, changes of the patch are separated from irrelevant changes. 
Third, we extract the test suite at the time of the patch commit, including failing test cases.\footnote{In considered commits, bug fixes are always committed together with originally failing test cases (which are passed after fixing the bugs). This is a rule in Apache development process \cite{apache2014}.} 
Fourth, we manually configure programs and test suites to examine whether bugs can be reproduced. 
Note that the reproducibility rate is very low due to the complexity of the projects Math and Lang.

\begin{table*}[!t]
\caption{The Evaluation Dataset of \nopol. It contains \numbug bugs related to buggy \ourif conditions and missing preconditions. }
\label{tab:bug}
\centering
\resizebox{1.01\textwidth}{!}{
\setlength\tabcolsep{0.2 ex}
\begin{tabular}{|c|c|ccc|c|c|c|c|lc|}

\hline 
\multirow{2}{*}{ \tabincell{c}{ Bug \\ type }} & \multirow{2}{*}{Project} & \multicolumn{3}{c|}{ Bug description }  & \multirow{2}{*}{\tabincell{c}{\scriptsize Executable \\ LoC }} &  \multirow{2}{*}{\#Classes}  &  \multirow{2}{*}{\#Methods}  &  \multirow{2}{*}{\tabincell{c}{ \#Unit \\ tests }}    &  \multicolumn{2}{c|}{ Buggy method }           \\ \cline{3-5}\cline{10-11}
 &   &   Index &   Commit ID\dag &   Bug ID\ddag &    & & & &      Description & \tiny Complexity      \\ \hline\hline
\multirow{16}{*}{ \rotatebox{90}{Buggy \ourif condition} }
 &  Math & CM1  & 141003& -        &    4611 &  153 &  947 &  363 & Returns a specified percentile from an array of real numbers &   7 \\
 &  &  CM2  &   141217  & -        &    5539 &  212 & 1390 &  543 & Returns an exact representation of the Binomial Coefficient  &   8 \\
 &  &  CM3  &   141473  & -        &    6633 &  191 & 1504 &  654 & Returns the natural logarithm of the factorial for a given value  &   3 \\
 &  &  CM4  &   159727  & -        &    7442 &  206 & 1640 &  704 & Computes the a polynomial spline function        &  5 \\
 &  &  CM5  &   735178  & Math-238 &   25034 &  468 & 3684 & 1502 &  Gets the greatest common divisor of two numbers & 12 \\
 &  &  CM6  &   791766  & Math-280 &   37918 &  632 & 5123 & 1982 & Finds two real values for a given univariate function   & 11 \\
 &  &  CM7  &   831510  & Math-309 &   38841 &  650 & 5275 & 2105 & Returns a random value from an Exponential distribution & 3  \\
 &  &  CM8  &   1368253 & Math-836 &   64709 &  975 & 7374 & 3982 & Converts a double value into a fraction       & 11 \\
 &  &  CM9  &   1413916 & Math-904 &   70317 & 1037 & 7978 & 4263 & Computes a power function of two real numbers & 38 \\
 &  &  CM10 &   1453271 & Math-939 &   79701 & 1158 & 9074 & 4827 & Checks whether a matrix has sufficient data to calculate covariance &   3 \\  
 & Lang & CL1 & 137231  & -        &   10367 &  156 & 2132 &  781 & Replaces a string with another one inside      &  4 \\
 &  &  CL2  &   137371  & -        &   11035 &  169 & 2240 &  793 & Removes a line break at the end of a string   &  4 \\
 &  &  CL3  &   137552  & -        &   12852 &  173 & 2579 &  994 & Gets a sub-string from the middle of a string from a given index & 5\\
 &  &  CL4  &   230921  & -        &   15818 &  215 & 3516 & 1437 & Finds the first matched string from the given index    & 10 \\
 &  &  CL5  &   825420  & Lang-535 &   17376 &   86 & 3744 & 1678 & Extracts the package name from a string       &  6 \\
 &  &  CL6  &   1075673 & Lang-428 &   18884 &  211 & 3918 & 1934 & Checks whether the char sequence only contains unicode letters &  4 \\
 \hline\hline
\multirow{6}{*}{ \rotatebox{90}{ \tabincell{c}{ Missing \\ precondition } } }
 &  Math &PM1 & 620221  & Math-191 &   16575 &  396 & 2803 & 1168 & Checks the status for calculating the sum of logarithms & 1 \\
 &  &  PM2  &  1035009  & -        &   44347 &  745 & 5536 & 2236 & Builds a message string from patterns and arguments &   3 \\     
 & Lang & PL1 & 504351  & Lang-315 &   17286 &  233 & 4056 & 1710 & Stops a process of timing   &   3 \\
 &  &  PL2  &   643953  & Lang-421 &   17780 &  240 & 4285 & 1829 & Erases a string with the Java style from the character stream &   19  \\
 &  &  PL3  &   655246  & Lang-419 &   18533 &  257 & 4443 & 1899 & Abbreviates a given string  &   9 \\
 &  &  PL4  &   1142389 & Lang-710 &   18974 &  218 & 3887 & 1877 & Counts and translates the code point from an XML numeric entity  &   19  \\
\hline\hline
\multicolumn{5}{|c|}{ Average }   & 25480.6 & 399.1 & 3960.4 & 1784.6 &   &  8.6  \\
\multicolumn{5}{|c|}{ Median }    & 17585.0 & 225.5 & 3818.5  & 1694.0    &   &  5.5  \\
\hline
       
\end{tabular}
}
\\
\ 

\tabfootnote{\dag \ A commit ID is an identifier that indicates the commit of the patch, in both SVN and the FishEye system. According to this commit, we can manually check relevant patched code and test cases. For instance, the commit of Bug CM1 can be found at {\tiny \url{https://fisheye6.atlassian.com/changelog/commons?cs=141003}.}} \\

\tabfootnote{\ddag \ For some bugs, bug IDs are not obviously identified in the bug tracking system. These bugs can be found in the version control system. For example, Apache projects previously used Bugzilla as a bug tracking system before moving to Jira. The Bugzilla system is not available anymore.}
\end{table*}

Table \ref{tab:bug} summarizes the \numbug bugs in two categories, i.e., bug types of buggy \ourif conditions and missing preconditions. We index these bugs according to their types and projects. A \textit{bug index} (Column 3) is named based on the following rule. Letters \textit{C} and \textit{P} indicate bugs with \buggyconditions and \preconditions, respectively; \textit{M} and \textit{L} are bugs from Math and Lang, respectively. For instance, CM1 refers to a bug with a \buggycondition in the project Math. We also record the number of executable Lines of Code (LoC, i.e., the number of lines that exclude empty lines and comment lines) for each source program (Column 6). 
Moreover, we show the number of classes, the number of methods in the buggy program, and the number of unit test cases (Columns 7-9). 
For each method that contains the buggy code, we describe the functionality of this method and record its Cyclomatic Complexity (Columns 10 and 11). 
The \textit{Cyclomatic Complexity} \cite{mccabe1976complexity} is the number of linearly independent paths through the source code of a method. This complexity indicates the testability of a method and the difficulty of understanding code by developers. 

As shown in Table \ref{tab:bug}, the dataset contains 16 bugs with buggy \ourif conditions and 6 bugs with missing preconditions. Among these bugs, 12 bugs are from Math and 10 bugs are from Lang. In average, a buggy program consists of 25.48K executable lines of code. The average complexity is 8.6; that is, a buggy method consists of 8.6 independent paths in average. Note that the method complexity of Bug PM1 is 1 since its buggy method contains only one \mycode{throw} statement (which misses a precondition); the method complexity of Bug CM9 is 38 and its buggy method contains 30 \ourif statements.

\subsection{Implementation Details}
\label{subsect:impl}

Our approach, \nopol, is implemented with Java 1.7 on top of Spoon 3.1.0.\footnote{Spoon 3.1.0, \url{http://spoon.gforge.inria.fr/}.} Spoon \cite{pawlak2015spoon} is a library for transforming and analyzing Java source code. It is used for angelic fix localization, instrumentation, and final patch synthesis in our work. Fault localization is implemented with GZoltar 0.0.10.\footnote{GZoltar 0.0.10, \url{http://gzoltar.com/}.} GZoltar \cite{CamposRPA12} is a fault localization library for ranking faulty statements. The SMT solver inside \nopol is Z3 4.3.2.\footnote{Z3, \url{http://github.com/Z3Prover/z3/}.} We generate SMT-LIB\footnote{SMT-LIB, \url{http://smt-lib.org/}.} files using jSMTLIB.\footnote{jSMTLIB, \url{http://sourceforge.net/projects/jsmtlib/}.} jSMTLIB \cite{cok2011jsmtlib} is a library for checking, manipulating, and translating SMT-LIB formatted problems. The test driver is JUnit 4.11. For future replication of the evaluation, the code of \nopol is available on GitHub \cite{nopolcode2014}. 

All experiments are run on a PC with an Intel Core i7 3.60 GHz CPU and a Debian 7.5 operating system. The maximum heap size of Java virtual machine was set to 2.50 GB.

\subsection{Main Research Questions}
\label{subsect:general-repair}

We present the general evaluation of \nopol on the dataset via answering six Research Questions (RQs). 

\medskip
\begin{mdframed}
RQ\refstepcounter{rqs}\arabic{rqs}\label{rq-fix}: Can \nopol fix real bugs in large-scale Java software?
\end{mdframed}

In test-suite based repair, a bug is \textit{fixed} if the patched program passes the whole test suite \cite{le2012genprog}. 
Table \ref{tab:patch} presents the evaluation of patches on \numbug bugs. 
Column 3 shows the buggy code (the condition for each bug with a \buggycondition and the statement for each bug with a \precondition).
Column 4 shows the patches that were manually-written by developers as found in the version control system: the updated condition for each bug with a \buggycondition and the added precondition for each bug with a \precondition. Column 5 presents the generated patches by \nopol. 
Column 6 is the result of our manual analysis of the correctness of the patches (will be explained in RQ\ref{rq-correct}).
Finally, Column 7 shows whether we had to modify existing test cases: ``A'' stands for additional test cases, ``T'' for transformed test cases, and ``D'' for deleted test cases.
The purpose of test case modification is to yield a correct repair (will be explained in RQ\ref{rq-testcase}).

\begin{table*}[!t]
\caption{Buggy code, manually-written patches, and generated patches for the bugs of the dataset.}
\label{tab:patch}
\centering
\resizebox{1.01\textwidth}{!}{
\setlength\tabcolsep{0.1 ex}

\renewcommand\arraystretch{1.5}
\begin{tabular}{|c|c|c|c|c|c|c|}

\hline 
\tabincell{c}{Bug \\ type} & \tabincell{c}{Bug \\ index} &  Buggy code &Patch written by developers & Patch generated by \nopol  & \scriptsize Correctness & \scriptsize \tabincell{c}{ Test case\\ modification}       \\   \hline\hline
\multirow{16}{*}{ \rotatebox{90}{Buggy \ourif condition} } & CM1 & \mycode{pos > n } & \mycode{pos >= n }  & \mycode{length <= fpos} & Correct  & A \ddag \\  \cline{2-7}

  & CM2 & \mycode{n <= 0 }  &  \mycode{n < 0 } & \mycode{if (n < MathUtils.ZS) \{ \ldots \ } \dag & Correct   & A T \\  \cline{2-7}

  & CM3 & \mycode{n <= 0 }  & \mycode{n < 0 }& \mycode{if (MathUtils.ZS != n) \{ \ldots }& Correct   & A T\\  \cline{2-7}
  & CM4 & \mycode{v < knots[0] || v >= knots[n]} & \mycode{v < knots[0] || v > knots[n]} & \scriptsize \texttt{if(v <= -1 || knots.length != v \&\& n < v +1)\{\ldots}&  Incorrect   & A\\ \cline{2-7}  
  & CM5 & \mycode{u * v == 0 }  & \mycode{u == 0 || v == 0 }& \mycode{if(u> MathUtils.NB \&\& u<= MathUtils.ZB} & Correct   & A T \\  
  &     &                       &                           & \mycode{|| v == MathUtils.ZB) \{ \ldots \ }       &           & \\ \cline{2-7} 

  & CM6 & \mycode{fa * fb >= 0.0 }  & \mycode{fa * fb > 0.0} & \mycode{if (-1 == b) \{ \ldots  }   & Incorrect   &  T D \\  \cline{2-7}
  & CM7 & \mycode{mean < 0 } & \mycode{mean <= 0 }& \mycode{mean <= 0.0}   & Correct   & A T D\\  \cline{2-7}
  & CM8 & \mycode{p2 > overflow || q2 > overflow}  & \mycode{FastMath.abs(p2) > overflow || }  & \mycode{-} (Timeout in SMT) & -  & - \\  
  &     &                                          & \mycode{FastMath.abs(q2) > overflow}      &  &   &  \\   \cline{2-7}
  & CM9 & \scriptsize \texttt{y >= TWO\_POWER\_52 || y <= -TWO\_POWER\_52} & \scriptsize \texttt{y >= TWO\_POWER\_53 || y <= -TWO\_POWER\_53}  & \mycode{-} (Timeout in test execution) &  -   & -\\  \cline{2-7}

 & CM10 & \mycode{nRows < 2 || nCols < 2 }  & \mycode{nRows < 2 || nCols < 1 } & \mycode{nRows < 2} & Correct & -\\ \cline{2-7} 

  & CL1 & \mycode{text == null }  & \mycode{text == null || repl == null ||}  & \mycode{if (with.length() == 0) \{ \ldots}  & Incorrect  & T D \\  
  &     &                         & \mycode{with == null || repl.length() == 0}  &                             &   &  \\  \cline{2-7}
  & CL2 & \mycode{lastIdx == 0 }  & \mycode{lastIdx <= 0 }  & \mycode{str.length() <= blanks.length()}    & Correct &  T\\  \cline{2-7}
  & CL3 & \mycode{pos > str.length()}  & \mycode{len < 0 || pos > str.length()}  & \mycode{len <= -1 || str.length() < pos}  & Correct   & T\\  \cline{2-7}
  & CL4 & \mycode{startIndex >= size}  & \mycode{substr == null || startIndex >= size} & \mycode{!(substr != null)||startIndex >= size} & Correct  & T D \\ \cline{2-7} 
  & CL5 & \mycode{className == null } & \scriptsize \texttt{className == null || className.length() == 0 }& \mycode{className.length() == 0 }    & Incorrect    & T \\  \cline{2-7}   
  & CL6 & \mycode{cs == null }  & \mycode{cs == null || cs.length() == 0 }  & \mycode{-} (Timeout due to a \mycode{null} pointer)  & - & T \\    
  \hline\hline
\multirow{6}{*}{\rotatebox{90}{\tabincell{c}{Missing \\ precondition }} } & PM1 & \scriptsize \texttt{throw new IllegalStateException("")} & \mycode{if (getN() > 0) \{ \ldots }  & \mycode{-} (No angelic value found) & -  & - \\    \cline{2-7}
  & PM2 & \mycode{sb.append(": ") } & \mycode{if (specific != null) \{ \ldots }  & \mycode{if (specific != null) \{ \ldots}    & Correct  & - \\    \cline{2-7}
  & PL1 & \scriptsize \texttt{stopTime $\leftarrow$ System.currentTimeMillis()} & \scriptsize \texttt{if (this.runningState == STATE\_RUNNING) \{ \ldots}& \mycode{if (stopTime < STATE\_RUNNING) \{ \ldots}   & Correct  & -\\    \cline{2-7}
  & PL2 & \mycode{out.write('\char`\\\char`\\') } & \mycode{if (escapeForwardSlash) \{ \ldots }& \mycode{if (escapeSingleQuote) \{ \ldots }   & Correct  & -\\    \cline{2-7}
  
  & PL3 & \mycode{lower $\leftarrow$ str.length() }  & \mycode{if (lower > str.length()) \{ \ldots } & \mycode{-} (Timeout in SMT)   & - & T \\    \cline{2-7} 
  & PL4 & \mycode{return 0 }  & \mycode{if (start == seqEnd) \{ \ldots }& \mycode{if (start == seqEnd) \{ \ldots  }    & Correct   & - \\   \hline 
\end{tabular}
}
\ 

\tabfootnote{\dag \ An added precondition is in a form of ``\texttt{if( ) \{ \ldots }'' to distinguish with an updated condition. }\\

\tabfootnote{\ddag \ For test case modification, ``A'' stands for additional test cases, ``T'' for transformed test cases, and ``D'' for deleted test cases.}\\ 
\end{table*}

As shown in Table \ref{tab:patch}, among \numbug bugs, \nopol can fix 17 bugs: 13 out of 16 bugs with \buggyconditions and 4 out of 6 bugs with \preconditions. Meanwhile, four out of five unfixed bugs relate to timeout. In our work, the execution time of \nopol is limited to up to five hours. We will empirically analyze the fixed bugs in Section \ref{subsect:case-study-repair} and explore the limitations of our approach as given by the five unfixed bugs in Section \ref{subsect:limitation}. 

Table \ref{tab:patch} also shows that patches generated by \nopol consist of both primitive values and object-oriented features. For the object-oriented features, two major types can be found in the generated patches: nullness checking (patches of Bugs CL4 and PM2) and the \mycode{length()} method of strings (patches of Bugs CL1, CL2, CL3, and CL5).

Note that six bugs (Bugs CM2, CM3, CM4, CM5, CM6, and CM10) with \buggyconditions are fixed by adding preconditions rather than updating conditions. One major reason is that a non-\ourif statement is ranked above the buggy \ourif statement during the fault localization; then \nopol adds a patch, i.e., a precondition to this non-\ourif statement. Hence, the condition inside the buggy \ourif statement cannot be updated. This shows that those two kinds of patches intrinsically relate to each other. To further understand this phenomenon, we have performed repair only in the mode of ``condition'' in \nopol: the six bugs could also be fixed via only updating \ourif conditions. 

\medskip
\begin{mdframed}
RQ\refstepcounter{rqs}\arabic{rqs}\label{rq-correct}: Are the synthesized patches as correct as the manual patches written by the developer?
\end{mdframed}

In practice, a patch should be more than making the test suite pass since test cases may not be enough for specifying program behaviors \cite{monperrus2014critical}, \cite{qi2015efficient}. 
In this paper, a generated patch is \textit{correct} if and only if the patch is functionally equivalent to the manually-written patch by developers.

For each synthesized patch, we have followed Qi et al. \cite{qi2015efficient} to perform a manual analysis of its correctness. 
The manual analysis consists of understanding the domain (e.g., the mathematical function under test for a bug in the project Math), understanding the role of the patch in the computation, and understanding the meaning of the test case as well as its assertions. 

As shown in Table \ref{tab:patch}, 13 out of 17 synthesized patches are as correct as the manual patches. Among these 13 correct patches, five patches (for Bugs CM1, CM2, CM3, CM5, and CM7) are generated based on not only the original test suite but also additional test cases. The reason is that the original test suite is too weak to drive the synthesis of a correct patch; then we had to manually write additional test cases (all additional test cases are publicly-available on the companion website \cite{nopol2014}\footnote{Additional test cases, \url{http://sachaproject.gforge.inria.fr/nopol/dataset/data/projects/math/}.}). This will be discussed in next research question. 

For four bugs (Bugs CM4, CM6, CL1, and CL5), we are not able to synthesize a correct patch. This will be further discussed in Section \ref{subsect:discussion-repair}. 

\medskip
\begin{mdframed}
RQ\refstepcounter{rqs}\arabic{rqs}\label{rq-testcase}: What is the root cause of test case modification?
\end{mdframed}

As shown in Table \ref{tab:patch}, some bugs are correctly repaired only after the test case modification (including test case addition, transformation, and deletion).  
The most important modification is test case addition. 
Six bugs (Bugs CM1, CM2, CM3, CM4, CM5, and CM7) with additional test cases correspond to too weak specifications. 
We manually added test cases for these bugs to improve the coverage of buggy statements. 
Without the additional test case, the synthesized patch is degenerated.
A case study of Bug CM1 (Section \ref{subsubsect:bug-cm1}) will further illustrate how additional test cases help to synthesize patches. 
Note that all additional test cases appear in the bugs, which are reported in the early stage of the project Math. One possible reason is that the early version of Math is not in test-driven development and the test suites are not well-designed. 

In test case transformation (Bugs CM2, CM3, CM5, CM6, CM7, CL1, CL2, CL3, CL4, CL5, CL6, and PL3), we simply break existing test cases into smaller ones, in order to have one assertion per test case. This is important for facilitating angelic fix localization since our implementation of \nopol has a limitation, which collects only one runtime trace for a test case (Section \ref{sect:angelic-fix-localization}). We note that such test case transformation can even be automated \cite{brefactoring}. 

The reason behind the four bugs with deleted test cases (Bugs CM6, CM7, CL1, and CL4) is accidental and not directly related to automatic repair: these deleted test cases are no longer compatible with the Java version and external libraries, which are used in \nopol. 

\medskip
\begin{mdframed}
RQ\refstepcounter{rqs}\arabic{rqs}\label{rq-specify}: How are the bugs in the dataset specified by test cases?
\end{mdframed}

To further understand the repair process of bugs by \nopol, Tables \ref{tab:patch-feature} and \ref{tab:nopatch-feature} show the detailed analysis for patched statements in 17 fixed bugs and buggy statements in five non-fixed bugs, respectively. Table~\ref{tab:patch-feature} gives the following information:
whether a synthesized patch is at the same location as the one written by the developer, 
the number of failing ($e_f$) and passing ($e_p$) test cases executing the patched statements,
the fault localization metrics (the rank of the patched statement and the total number of suspicious statements), the overall execution time of \nopol, and the SMT level (see Section \ref{sec:smt-level}).
In Table \ref{tab:nopatch-feature}, $e_f$ and $e_p$ denote the number of failing and passing test cases executing the buggy statements while the rank of the buggy statement is listed.

Tables \ref{tab:patch-feature} and \ref{tab:nopatch-feature} show the numbers of failing ($e_f$) and passing ($e_p$) test cases that execute the patched or buggy statements. Such numbers reflect to which extent the statement under repair is specified by test cases.
As shown in Table \ref{tab:patch-feature}, the average of $e_f$ and $e_p$ are 1.8 and 7.1 for 17 bugs with synthesized patches. In Table \ref{tab:nopatch-feature}, the average of $e_f$ and $e_p$ are 4.4 and 27.4 for five bugs without patches. 

For all 22 bugs under evaluation, only one bug has a large number of failing test cases  ($e_f\geq 10$): Bug PL3 with $e_f=18$. For this bug, although the buggy statement is ranked at the first place, \nopol fails in synthesizing the patch. 
This failure is caused by an incorrectly identified output of a precondition. 
Section \ref{subsubsect:limit-pl3} will explain the reason behind this failure.

\begin{table}[!t]
\caption{Analysis of the 17 fixed bugs (patched statement).}
\label{tab:patch-feature}
\centering
\resizebox{0.5\textwidth}{!}{
\setlength\tabcolsep{0.2 ex}
\begin{tabular}{|c|c|c|cc|cc|c|c|}

\hline 
\multirow{2}{*}{ \tabincell{c}{ Bug \\ type }}  & \multirow{2}{*}{ \tabincell{c}{ Bug \\ index}} & \multirow{2}{*}{\scriptsize \tabincell{c}{ Patch \\location }} &  \multicolumn{2}{c|}{\scriptsize \#Test cases }  & \multirow{2}{*}{\scriptsize \tabincell{c}{\textbf{Patched} \\ statement rank }}  & \multirow{2}{*}{ \scriptsize \tabincell{c}{  \#Suspicious \\statements \ddag}} &  \multirow{2}{*}{\tiny \tabincell{c}{Execution time\\ (seconds)}}  &  \multirow{2}{*}{ \tabincell{c}{SMT \\ level}} \\ \cline{4-5}
  & & & $e_f$\dag & $e_p$\dag  & & & & \\
\hline\hline

\multirow{13}{*}{ \rotatebox{90}{Buggy \ourif condition} } & CM1 & \scriptsize  Same as dev. & 1 & 4 & 57 & 203  &  12  & 2  \\  [0.5ex]
  & CM2 & Different & 1 & 2 & 179 & 559 &  11  & 2 \\
  & CM3 & Different & 1 & 2 & 26 & 35  &  10 & 2 \\
  & CM4 & Different & 6 & 2 & 24  & 114 &  13  & 3  \\  [0.5ex]
  & CM5 & Different & 4 & 4 & 3 & 60 &  43  & 3  \\  [0.5ex]
  & CM6 & Different & 2 & 0 & 17  & 254 &  41  & 2  \\
  & CM7 & \scriptsize  Same as dev. & 3 & 13 & 143  & 155  &  51  & 2  \\  [0.5ex]
  & CM10  & \scriptsize  Same as dev. & 1 & 3 & 89 & 102 &  21  & 1  \\
  & CL1 & Different & 2 & 4 & 2 & 8 &  32  & 2 \\  [0.5ex]
  & CL2 & \scriptsize  Same as dev. & 1 & 9 & 2 & 3 &  6 & 2  \\  [0.5ex]
  & CL3 & \scriptsize  Same as dev. & 2 & 10 & 4 & 5 &  7  & 3  \\  [0.5ex]
  & CL4 & \scriptsize  Same as dev. & 2 & 23 & 1 & 20  &  10  & 3  \\  [0.5ex]
  & CL5 & \scriptsize  Same as dev. & 1 & 36 & 1 & 2 &  37  & 1  \\  
  \hline \hline 
\multirow{4}{*}{ \rotatebox{90}{\scriptsize \tabincell{c}{Missing\\ precondition}}}   & PM2 & \scriptsize  Same as dev. & 1 & 2 & 1 & 12  &  84  & 1  \\  [0.5ex]
  & PL1 & \scriptsize  Same as dev. & 1 & 4 & 6 & 22  &  32  & 2  \\  [0.5ex]
  & PL2 & \scriptsize  Same as dev. & 1 & 1 & 1 & 21  &  6 & 1  \\  [0.5ex]
  & PL4 & \scriptsize  Same as dev. & 1 & 1 & 1 & 25  &  6 & 2  \\ 
  \hline\hline
\multicolumn{2}{|c|}{  Median } &  \multirow{2}{*}{}   & 1  & 4  & 4 & 25 &  13 & 2\\  [0.5ex]
\multicolumn{2}{|c|}{  Average } &                     & 1.8  & 7.1 & 32.8 & 94.1   &  24.8 & 2 \\

\hline
\end{tabular}
}
\ 

\tabfootnote{\dag \ $e_f$ and $e_p$ denote the number of failing and passing test cases that execute the \textbf{patched} statement.}\\

\tabfootnote{\ddag \ \#Suspicious statements denotes the number of statements whose suspiciousness scores by fault localization are over zero.}\\

\end{table}

\begin{table}[!t]
\caption{Analysis of the 5 non-fixed bugs (buggy statement).}
\label{tab:nopatch-feature}
\centering
\resizebox{0.45\textwidth}{!}{
\setlength\tabcolsep{0.3 ex}
\begin{tabular}{|c|c|cc|cc|c|}

\hline 
\multirow{2}{*}{ \tabincell{c}{ Bug \\ type }}  & \multirow{2}{*}{ \tabincell{c}{ Bug \\ index}} & \multicolumn{2}{c|}{\scriptsize \#Test cases }  & \multirow{2}{*}{\scriptsize \tabincell{c}{\textbf{Buggy} \\statement rank }}  & \multirow{2}{*}{ \scriptsize \tabincell{c}{  \#Suspicious \\statements \ddag}} &  \multirow{2}{*}{\scriptsize \tabincell{c}{Execution time\\ (seconds) }}   \\ \cline{3-4}
  & & $e_f$\dag & $e_p$\dag  & & & \\

\hline\hline
\multirow{3}{*}{\scriptsize \tabincell{c}{Buggy \ourif\\ condition}}  
  & CM8 & 1 & 51 & 21  & 77  &  -  \\  [0.5ex]
  & CM9 & 1 & 73 & 1203 & 1606 &  -  \\  [0.5ex]
  & CL6 & 1 & 10 & 4 & 4 &  -   \\    
  \hline\hline

\multirow{2}{*}{\scriptsize \tabincell{c}{Missing \\precondition}} 
  & PM1 & 1 & 0 & 52 & 132 &  37  \\  [0.5ex]
  & PL3 & 18  & 3 & 1 & 16  & - \\  
  \hline\hline

\multicolumn{2}{|c|}{  Median }  & 1  & 10 & 21 & 77 &  -  \\  [0.5ex]
\multicolumn{2}{|c|}{  Average } & 4.4  & 27.4 & 256.2 & 367    &  -  \\
\hline

\end{tabular}
}
\ 

\tabfootnote{\dag \ $e_f$ and $e_p$ denote the number of failing and passing test cases that execute the \textbf{buggy} statement.}\\

\tabfootnote{\ddag \ \#Suspicious statements denotes the number of statements whose suspiciousness scores by fault localization are over zero.}\\

\end{table}

\medskip
\begin{mdframed}
RQ\refstepcounter{rqs}\arabic{rqs}: Where are the synthesized patches localized? How long is the repair process? 
\end{mdframed}

A patch could be localized in a different location from the patch which is manually-written by developers. We present the details for patch locations for all the 17 patched bugs in Table \ref{tab:patch-feature}. 
For 11 out of 17 fixed bugs, the locations of patched statements (i.e., locations of fixes) are exactly the same as those of the buggy ones. For the other six bugs, i.e., Bugs CM2, CM3, CM4, CM5, CM6, and CL1, \nopol generates patches by adding new preconditions rather than updating existing conditions, as mentioned in Table \ref{tab:patch}. 

For 17 fixed bugs in Table \ref{tab:patch-feature}, the average execution time of repairing one bug is 24.8 seconds while for five non-fixed bugs in Table \ref{tab:nopatch-feature}, four bugs are run out of time and the other one spends 37 seconds. The execution time of all the 22 bugs ranges from 6 to 84 seconds. We consider that such execution time, i.e., fixing one bug within 90 seconds, is acceptable. 

In practice, if applying \nopol to a buggy program, we can directly set a timeout, e.g., 90 seconds (over 84 seconds as shown in Table \ref{tab:patch-feature}) or a longer timeout like five hours in our experiment. Then for any kind of buggy program (without knowing whether the bug is with a buggy condition or a missing precondition), \nopol will synthesize a patch, if it finds any. Then a human developer can check whether this patch is correct from the user perspective. 

\medskip
\begin{mdframed}
RQ\refstepcounter{rqs}\arabic{rqs}:  How effective is fault localization for the bugs in the dataset?
\end{mdframed}

Fault localization is an important step in our repair approach. As shown in Table \ref{tab:patch-feature}, for patched statements in 17 fixed bugs, the average fault localization rank is 32.8. In four out of 17 bugs (Bugs CM1, CM2, CM7, and CM10), patched statements are ranked over 50. This fact indicates that there is room for improving the fault localization techniques. Among the five unfixed bugs, the buggy statements of Bugs CM9 and PM1 are ranked over 50. Section \ref{subsect:discussion-fault} will further compare the effectiveness of six fault localization techniques on 22 bugs. 

Note that in Tables \ref{tab:patch-feature} and \ref{tab:nopatch-feature}, Bugs CM6 and PM1 have no passing test cases. Bug PM1 cannot be fixed by our approach while Bug CM6 can be still fixed because the two failing test cases give a non-trivial input-output specification.
The reason behind the unfixed Bug PM1 is not $e_p=0$, but the multiple executions of the buggy code by one test case. This reason will be discussed in Section \ref{subsubsect:limit-pm1}.

\subsection{Case Studies for Fixed Bugs}
\label{subsect:case-study-repair}

We conduct four case studies to show how \nopol fixes bugs with \buggyconditions and \preconditions. These bugs are selected because they highlight different facets of the repair process. The patch of Bug PL4 (Section \ref{subsubsect:bug-pl4}) is syntactically the same as the manually-written one; 
the patch of Bug CL4 (Section \ref{subsubsect:bug-cl4}) is as correct as the manually-written patch (beyond passing the test suite); 
the patch of Bug CM2 (Section \ref{subsubsect:bug-cm2}) is correct by adding a precondition rather than updating the buggy condition, as written by developers; 
and the patch of Bug CM1 (Section~\ref{subsubsect:bug-cm1}) is correct, but its patch requires an additional test case.

\subsubsection{Case Study 1, Bug PL4}
\label{subsubsect:bug-pl4}

\nopol can generate the same patches for three out of \numbug bugs as the manually-written ones. 
We take Bug PL4 as an example to show how the same patch is generated. This bug is fixed by adding a precondition. Fig. \ref{fig:bug-pl4} shows a method \mycode{translate()} at Line 1 and the buggy method \mycode{translateInner()} at Line 9 of Bug PL4. The method \mycode{translate()} is expected to translate a term in the regular expression of \mycode{\&\#[xX]?\char`\\d+;?} into codepoints, e.g., translating the term \mycode{"\&\#x30"} into \mycode{"\char`\\u0030"}.

To convert from \mycode{input} to codepoints, the characters in \mycode{input} are traversed one by one. Note that for a string ending with \mycode{"\&\#x"}, no codepoint is returned. Lines 15 to 20 in Fig. \ref{fig:bug-pl4} implement this functionality. However, the implementation at Lines 17 to 19 ignores a term in a feasible form of \mycode{"\&\#[xX]\char`\\d+"}, e.g., a string like \mycode{"\&\#x30"}. A precondition should be added to detect this feasible form, i.e., the comment at Line 18 of \mycode{start == seqEnd}.

The buggy code at Line 19 is executed by one passing test case and one failing test case. Table \ref{tab:test-pl4} shows these two test cases. For the passing test case, the behavior of the method is expected not to change the variable \mycode{input} while for the failing test case, the \mycode{input} is expected to be converted. In the passing test case, the value of the precondition of the statement at Line 19 is expected to be \mycode{true}, i.e., both \mycode{start} and \mycode{seqEnd} equal to \mycode{8}, while in the failing test case, the condition is expected to be \mycode{false}, i.e., \mycode{start} and \mycode{seqEnd} are \mycode{8} and \mycode{19}, respectively. The \mycode{false} value is the angelic value for a missing precondition.

\begin{figure}[!t]
\centering
\noindent\begin{minipage}{0.4\textwidth}
\begin{lstlisting}[numbers=left]
String translate(CharSequence input, int index) {
  int consumed = translateInner(input, index);
  if(consumed == 0)
    ... //  Return the original input value
  else
    ... //  Translate code points
}

int translateInner(CharSequence input, int index) {
  int seqEnd = input.length();
  ...
  int start = index + 2;
  boolean isHex = false;
  char firstChar = input.charAt(start);
  if(firstChar == 'x' || firstChar == 'X') {
    start++;
    isHex = true;
//  FIX: if(start == seqEnd)    
    return 0;
  }
  int end = start;
  //Traverse the input and parse into codepoints
  while(end < seqEnd && ... )
  ...  
}
\end{lstlisting}

\end{minipage}
\caption{Code snippet of Bug PL4. The manually-written patch is shown in the \mycode{FIX} comment at Line 18. Note that the original method \mycode{translate} consists of three overloaded methods; for the sake of simplification, we use two methods \mycode{translate} and \mycode{translateInner} instead.}
\label{fig:bug-pl4}
\end{figure}

\begin{table}[!t]
\centering
\caption{Sample of test cases for Bug PL4}
\label{tab:test-pl4}

\resizebox{0.51\textwidth}{!}{
\setlength\tabcolsep{0.2 ex}
\begin{tabular}{|cc|cc|c|}

\hline
\multicolumn{2}{|c|}{Input} & \multicolumn{2}{c|}{Output, \mycode{translate(input)}} & \multirow{2}{*}{ \tabincell{c}{ Test \\ result} } \\ 
\cline{1-4}
 \mycode{input} & \mycode{index} &  Expected &  Observed & \\
\hline \hline
\mycode{"Test \&\#x"} & 5 &  \mycode{"Test \&\#x"} & \mycode{"Test \&\#x"}  & Pass\\
\mycode{"Test \&\#x30 not test"} & 5 & \mycode{"Test \char`\\u0030 not test"} & \mycode{"Test \&\#x30 not test"} & Fail\\

\hline
\end{tabular}
}
\end{table}

According to those expected precondition values for test cases, \nopol generates a patch via adding a precondition, i.e., \mycode{start == seqEnd}, which is exactly the same as the manually-written patch by developers. Besides the patch of Bug PL4, patches of Bugs CM7 and PM2 are also syntactically the same as the patch written by developers, among \numbug bugs in our dataset.

\subsubsection{Case Study 2, Bug CL4}
\label{subsubsect:bug-cl4}

For several bugs, \nopol generates patches that are literally different from the manually-written patches, but these generated patches are as correct as manually-written patches. In this section, we present a case study where \nopol synthesizes a correct patch for a bug with a \buggycondition. 
Bug CL4 in Lang fails to find the index of a matched string in a string builder. Fig. \ref{fig:bug-cl4} presents the buggy method of Bug CL4: to return the first index of \mycode{substr} in a parent string builder from a given basic index \mycode{startIndex}. The condition at Line 4 contains a mistake of \mycode{startIndex >= size}, which omits checking whether \mycode{substr == null}. A variable \mycode{size} is defined as the length of the parent string builder. The manually-written fix is shown at Line 3.

The buggy code at Line 4 in Fig. \ref{fig:bug-cl4} is executed by 23 passing test cases and two failing test cases. One of the passing test cases and two failing test cases are shown in Table \ref{tab:test-cl4}. For the passing test case, a value \mycode{-1} is expected because no matched string is found. For the two failing test cases, each input \mycode{substr} is a \mycode{null} value, which is also expected to return a non-found index \mycode{-1}. This requires the checking of \mycode{null} to avoid \mycode{NullPointerException}, i.e., the condition at Line 3.

For the passing test case in Table \ref{tab:test-cl4}, the condition at Line 4 is \mycode{false}. For the two failing test cases, \nopol extracts the angelic value \mycode{true} to make both failing test cases pass. According to these condition values, a patch of \mycode{! (substr != null) || startIndex >= size} can be synthesized. 
This synthesized patch is equivalent to \mycode{substr == null || startIndex >= size}, which is correct.
The resulted expression based on the solution to the SMT does not weaken the repairability of synthesized patches.
A recent method for finding simplified patches, proposed by Mechtaev et al. \cite{mechtaev2015directfix}, could be used to avoid such a redundant expression. 

\iffalse
For the passing test case, the condition at Line 4 is \mycode{false}. For the two failing test cases, \nopol extracts the angelic value \mycode{true} to make both failing test cases pass. According to these condition values, a patch of \mycode{substr == null || startIndex == size} can be synthesized. 

However, this patch is different from the manually-written one at Line 3. The difference is that the generated patch omits the comparison of \mycode{startIndex > size}. But the following code at Lines 9 to 20 in Fig. \ref{fig:bug-cl4} shows that the checking of \mycode{startIndex > size} is not necessary. The \mycode{for} loop at Line 13 starts the traversal of the string builder. Once \mycode{startIndex > size} holds, the loop at Line 13 does not perform and the execution of the code jumps to Line 24. The code at Line 24 returns the same value as the expected values in the failing test cases. Hence the implementation of checking \mycode{startIndex > size} already exists in the code. Such existing implementation inside the source code makes the equivalent patches possible. 
\fi

\subsubsection{Case Study 3, Bug CM2} 
\label{subsubsect:bug-cm2}

In this section, we present Bug CM2, a correctly patched bug via adding a precondition, rather than updating an existing condition, as written by developers. 
The buggy method in Bug CM2 is to calculate the value of Binomial Coefficient by choosing \mycode{k}-element subsets from an \mycode{n}-element set. Fig. \ref{fig:bug-cm2} presents the buggy method. The input number of elements \mycode{n} should be no less than zero. But the condition at Line 4 reads \mycode{n <= 0} instead of \mycode{n < 0}. The manually-written patch by developers is in the \mycode{FIX} comment at Line 4.

\begin{figure}[!t]
\centering
\noindent\begin{minipage}{0.4\textwidth}
\begin{lstlisting}[numbers=left]
int indexOf(String substr, int startIndex) {
  startIndex = (startIndex < 0 ? 0 : startIndex);
// FIX: if (substr == null || startIndex >= size) {
  if (startIndex >= size) {
    return -1;
  }
  int strLen = substr.length();
  if (strLen > 0 && strLen <= size) {
    if (strLen == 1) 
      return indexOf(substr.charAt(0), startIndex);
    char[] thisBuf = buffer;
    outer:
    for (int i = startIndex; i < thisBuf.length
            - strLen; i++) {
      for (int j = 0; j < strLen; j++) {
        if (substr.charAt(j) != thisBuf[i + j]) 
          continue outer;
      }
      return i;
    }
  } else if (strLen == 0) {
    return 0;
  }
  return -1;
}
\end{lstlisting}

\end{minipage}
\caption{Code snippet of Bug CL4. The manually-written patch is shown in the \mycode{FIX} comment at Line 3, which updates the buggy \ourif condition at Line 4.}
\label{fig:bug-cl4}
\end{figure}

\begin{table}[!t]
\centering
\caption{Sample of test cases for Bug CL4}
\label{tab:test-cl4}

\resizebox{0.51\textwidth}{!}{
\setlength\tabcolsep{0.2 ex}
\begin{tabular}{|ccc|cc|c|}

\hline
\multicolumn{3}{|c|}{Input} &  \multicolumn{2}{c|}{\tiny Output, \texttt{indexOf(substr, startIndex)}} & \multirow{2}{*}{ \tabincell{c}{ Test \\ result} }  \\ 
\cline{1-5}
\mycode{parent} & \mycode{substr} & \mycode{startIndex} &  Expected &  Observed & \\
\hline\hline
abab & z & 2 & -1 & -1 & Pass \\
abab & (String) null & 0  & -1 & \tiny \texttt{NullPointerException} & Fail  \\
xyzabc & (String) null & 2 & -1 &  \tiny \texttt{NullPointerException} & Fail  \\
\hline
\end{tabular}
}
\end{table}

The buggy code at Line 4 in Fig. \ref{fig:bug-cm2} is executed by two passing test cases and one failing test case. Table \ref{tab:test-cm2} shows one passing test case and one failing test case. For the passing test case, an expected exception is observed; for the failing test case, an \mycode{IllegalArgumentException} is thrown rather than an expected value. 

To fix this bug, \nopol generates a patch via adding a missing precondition \mycode{n < MathUtils.ZS} to the statement at Line 5, where \mycode{MathUtils.ZS} is a constant equal to \mycode{0}. 
Then this statement owns two embedded preconditions, i.e., \mycode{n <= 0} and \mycode{n < 0}. 
Hence, the generated patch is equivalent to the manually-written patch, i.e., updating the condition at Line 4 from \mycode{n <= 0} to \mycode{n < 0}. 
The reason of adding a precondition instead of updating the original condition is that the statement at Line 5 is ranked prior to the statement at Line 4. This has been explained in Section \ref{subsect:general-repair}.  
Consequently, the generated patch of Bug CM2 is correct and syntactically equivalent to the manually-written patch.

\begin{figure}[!t]
\centering
\noindent\begin{minipage}{0.4\textwidth}
\begin{lstlisting}[numbers=left]
long binomialCoefficient(int n, int k) {
  if (n < k) 
    throw new IllegalArgumentException(...);
  if (n <= 0)    //  FIX: if (n < 0)  
    throw new IllegalArgumentException(...);
  if ((n == k) || (k == 0)) 
    return 1;
  if ((k == 1) || (k == n - 1)) 
    return n;
  long result = Math.round(
      binomialCoefficientDouble(n, k));
  if (result == Long.MAX_VALUE) 
    throw new ArithmeticException(...);
  return result;
}
\end{lstlisting}

\end{minipage}
\caption{Code snippet of Bug CM2. The manually-written patch is shown in the \mycode{FIX} comment at Line 4.}
\label{fig:bug-cm2}
\end{figure}

\begin{table}[!t]
\centering
\caption{Sample of test cases for Bug CM2}
\label{tab:test-cm2}

\resizebox{0.47\textwidth}{!}{
\setlength\tabcolsep{0.3 ex}
\begin{tabular}{|cc|cc|c|}

\hline
\multicolumn{2}{|c|}{Input} & \multicolumn{2}{c|}{Output, \mycode{\scriptsize{binomialCoefficient(n,k)}} } & \multirow{2}{*}{ \tabincell{c}{ Test \\ result} } \\ 
\cline{1-4}
 \mycode{n} & \mycode{k} &  Expected&  Observed& \\
\hline \hline
 -1 & -1 & Exception & Exception & Pass\\
 0 & 0 &  1 & Exception  & Fail\\
\hline
\end{tabular}
}
\end{table}

\subsubsection{Case Study 4, Bug CM1}
\label{subsubsect:bug-cm1}

Insufficient test cases lead to trivial patch generation. 
We present Bug CM1 in Math, with a \buggycondition. This bug cannot be correctly patched with the original test suite due to the lack of test cases. In our work, we add two test cases to support the patch generation. 
Fig. \ref{fig:bug-cm1} presents the buggy source code in the method \mycode{evaluate()} of Bug CM1. This method returns an estimate of the percentile \mycode{p} of the values stored in the array \mycode{values}. 

According to the API document, the algorithm of \mycode{evaluate()} is implemented as follows. 
Let \mycode{n} be the length of the (sorted) array. The algorithm computes the estimated percentile position \mycode{pos = p * (n + 1) / 100} and the difference \mycode{dif} between \mycode{pos} and \mycode{floor(pos)}. If \mycode{pos >= n}, then the algorithm returns the largest element in the array; otherwise the algorithm returns the final calculation of percentile. Thus, the condition at Line 15 in Fig. \ref{fig:bug-cm1} contains a bug, which should be corrected as \mycode{pos >= n}. 

As shown in Table \ref{tab:patch-feature}, this bug is executed by four passing test cases and one failing test case. Table \ref{tab:test-cm1} shows one of the four passing test cases and the failing test case. In the failing test case, an \mycode{ArrayIndexOutOfBounds} exception is thrown at Line 16. For the passing test case, the value of the condition at Line 15 is equal to the value of the existing condition \mycode{pos > n}, i.e., \mycode{true}; for the failing test case, setting the condition to be \mycode{true} makes the failing test case pass; that is, the angelic value for the failing test case is also \mycode{true}. Thus, according to these two test cases, the generated patch should make the condition be \mycode{true} to pass both test cases.

\begin{figure}[!t]
\centering
\noindent\begin{minipage}{0.4\textwidth}
\begin{lstlisting}[numbers=left]
double evaluate(double[] values, double p) {
  ...
  int length = values.length;
  double n = length;
  ...
  double pos = p * (n + 1) / 100;
  double fpos = Math.floor(pos);
  int intPos = (int) fpos;
  double dif = pos - fpos;
  double[] sorted = new double[n];
  System.arraycopy(values, 0, sorted, 0, n);
  Arrays.sort(sorted);
  if (pos < 1) 
      return sorted[0];   
  if (pos > n)     //  FIX: if (pos >= n) 
      return sorted[n - 1];    
  double lower = sorted[intPos - 1];
  double upper = sorted[intPos];
  return lower + dif * (upper - lower);
}
\end{lstlisting}

\end{minipage}
\caption{Code snippet of Bug CM1. The manually-written patch is shown in the \mycode{FIX} comment at Line 15.}
\label{fig:bug-cm1}
\end{figure}

\begin{table}[!t]
\centering
\caption{Two original test cases and one additional test case for Bug CM1}
\label{tab:test-cm1}

\resizebox{0.48\textwidth}{!}{
\setlength\tabcolsep{0.3 ex}
\begin{tabular}{|cc|cc|c|}

\hline
\multicolumn{2}{|c|}{Input} & \multicolumn{2}{c|}{Output, \mycode{\scriptsize evaluate(values,p)}} & \multirow{2}{*}{ \tabincell{c}{ Test \\ result} } \\ 
\cline{1-4}
 \mycode{values} & \mycode{p} &  Expected&  Observed& \\
\hline \hline
 \multicolumn{5}{|c|}{Two original test cases} \\ \hline
 \{0,1\} & 25 & 0.0 & 0.0 & Pass\\
 \{1,2,3\} & 75 &  3.0 & Exception  & Fail\\
\hline \hline
 \multicolumn{5}{|c|}{Two additional test cases} \\ \hline 
 \{1,2,3\} & 100 &  3.0 & 3.0 & Pass \\
\hline
\end{tabular}
}
\end{table}

\begin{table*}[!t]
\caption{Summary of five limitations.}
\label{tab:limitation}
\centering
\resizebox{1\textwidth}{!}{
\setlength\tabcolsep{0.7 ex}
\begin{tabular}{|c|lp{45 ex}p{62 ex}|}

\hline
Bug index & Root cause & Result of repair & Reason for the unfixed bug \\ \hline\hline
PM1 & Angelic fix localization & No angelic value found. Termination before runtime trace collection and patch synthesis & One failing test case executes the missing precondition for more than once. \\  [0.6 ex]
CM9 & Angelic fix localization & Timeout during test suite execution &  An infinite loop is introduced during the trial of angelic values.  \\ [0.6 ex]
PL3 & Runtime trace collection & Timeout during SMT solving & The expected value of a precondition is incorrectly identified.  \\ [0.6 ex]
CM8 & Patch synthesis & Timeout during SMT solving& A method call with parameters is not handled by SMT.  \\ [0.6 ex]
CL6 & Patch synthesis & Timeout during SMT solving & A method of a \mycode{null} object yields an undefined value for SMT. \\
\hline

\hline
\end{tabular}
}
\end{table*}

With the original test suite, \nopol generates a patch as \mycode{sorted.length <= intPos}, which passes all test cases. This patch is incorrect. 
To obtain a correct patch (the one shown in Table \ref{tab:patch}), we add one test case of \mycode{values $\leftarrow$ \{1,2,3\}}, \mycode{p $\leftarrow$ 100}, as shown in Table~\ref{tab:test-cm1}. Then the expected value of \mycode{evaluate()} is \mycode{3.0}. After running \nopol, a patch of \mycode{length <= fpos}, which is different from the manually-written one, i.e., \mycode{pos >= n}. However, from the source code at Line 7, \mycode{fpos} is the \mycode{floor()} value of \mycode{pos}, i.e., \mycode{fpos} is the largest integer that no more than \mycode{pos}. That is, \mycode{fpos <= pos}. Meanwhile, \mycode{n == length} holds according to Line 4. As a result, the generated patch \mycode{length <= fpos} implies the manually-written one, i.e., \mycode{pos >= n}. We can conclude that \nopol can generate a correct patch for this bug by adding one test case.

\subsection{Limitations}
\label{subsect:limitation}  

As shown in Section \ref{subsect:general-repair}, we have collected five bugs that reveal five different limitations of \nopol. Table \ref{tab:limitation} lists these five bugs in details. We analyze the related limitations in this section.

\subsubsection{No Angelic Value Found}
\label{subsubsect:limit-pm1}

In our work, for a buggy \ourif condition, we use angelic fix localization to flip the boolean value of conditions for failing test cases. 
For Bug PM1, no angelic value is found as shown in Table \ref{tab:patch}. 
The reason is that both \mycode{then} and \mycode{else} branches of the \ourif condition are executed by one failing test case. Hence, no single angelic value (\texttt{true} or \texttt{false}) can enable the test case to pass. As discussed in Section \ref{subsubsect:search-space}, the search space of a sequence of angelic values is exponential and hence discarded in our implementation of \nopol.

To mitigate this limitation, a straightforward solution is to discard the failing test case, which leads to no angelic values (keeping the remaining failing ones). However, this may decrease the quality of the generated patch due to the missing test data and oracles. Another potential solution is to refactor test cases into small snippets, each of which covers only \mycode{then} or \mycode{else} branches \cite{brefactoring}. A recent proposed technique SPR \cite{DBLP:conf/sigsoft/LongR15} could help \nopol to enhance its processing of sequential angelic values.

\subsubsection{Performance Bugs Caused by Angelic Values}

\nopol identifies angelic values as the input of patch synthesis. In the process of angelic fix localization, all failing test cases are executed to detect conditional values that make failing test cases pass (see Section \ref{sect:angelic-fix-localization}). However, sometimes the trial of angelic fix localization (forcing to \mycode{true} or \mycode{false}) may result in a performance bug. In our work, Bug CM9 cannot be fixed due to this reason, i.e., an infinite loop caused by angelic fix localization.

A potential solution to this issue is to set a maximum execution time to avoid the influence of performance bugs. But a maximum execution time of test cases may be hard to be determined according to different test cases. For instance, the average execution time of test cases in Math 3.0 is much longer than that in Math 2.0. We leave the setting of maximum execution time as one piece of future work.

\subsubsection{Incorrectly Identified Output of a Precondition}
\label{subsubsect:limit-pl3}

As mentioned in Section \ref{subsubsect:outcome}, the expected output of a missing precondition is set to be \mycode{true} for a passing test case and is set to be \mycode{false} for a failing one. The underlying assumption for a passing test case is that the \mycode{true} value keeps the existing program behavior. However, it is possible that given a statement, both \mycode{true} and \mycode{false} values can make a test case pass. 
In these cases, synthesis may not work for bugs with missing preconditions.

This is what happens to Bug PL3. Fig. \ref{fig:bug-pl3} shows a code snippet of Bug PL3. The manually-written patch is a precondition at Line 3; Table \ref{tab:test-pl3} shows one passing test case and one failing test case. 
Based on the angelic fix localization in Algorithm \ref{alg:missing-pred-angelic-fix-localization}, the expected precondition values of all passing test cases are set to be \mycode{true}. However, in the manually-written patch, the precondition value by the passing test case in Table~\ref{tab:test-pl3} is \mycode{false}, i.e., \mycode{lower > str.length()} where \mycode{lower} is \mycode{0} and \mycode{str.length()} is \mycode{10}. 
Thus, it is impossible to generate a patch like the manually-written one, due to a conflict in the input-output specification. Consequently, in the phase of patch synthesis, the SMT solver executes with timeout.

The example in Bug PL3 implies that for some bugs, the assumption (i.e., a missing precondition is expected to be \mycode{true} for passing test cases) can be violated. For Bug PL3, we have temporarily removed this assumption and only used the failing test cases to synthesize a patch. The resulting patch is \mycode{if(lower >= str.length()) lower = str.length()}, which has the same program behavior as the manually-written patch, i.e., \mycode{lower > str.length()}. In \nopol, we are conservative and assume that the expected value of a precondition by passing test cases is \mycode{true} (in Section~\ref{subsubsect:precondition}).

\subsubsection{Complex Patches using Method Calls with Parameters}

In our work, we support the synthesis of conditions that call unary methods (without parameters). However, our approach cannot generate a patch if a method with parameters has to appear in a condition. For example, for Bug CM8, the patch that is written by developers contains a method \mycode{abs(x)} for computing the absolute value. Our approach cannot provide such kinds of patches because methods with parameters cannot be directly encoded in SMT. Then the lack of information of method calls leads to the timeout of an SMT solver.

\begin{figure}[!t]
\centering
\noindent\begin{minipage}{0.4\textwidth}
\begin{lstlisting}[numbers=left]
String abbreviate(String str, int lower, int upper){
  ...
  //  FIX: if (lower > str.length()) 
  lower = str.length();    

  if (upper == -1 || upper > str.length()) 
    upper = str.length();
  if (upper < lower) 
    upper = lower;
  StringBuffer result = new StringBuffer();
  int index = StringUtils.indexOf(str, " ", lower);
  if (index == -1) 
    result.append(str.substring(0, upper));    
  else ...
  return result.toString();
}
\end{lstlisting}

\end{minipage}
\caption{Code snippet of Bug PL3. The manually-written patch is shown in the \mycode{FIX} comment at Line 3.}
\label{fig:bug-pl3}
\end{figure}

\begin{table}[!t]
\centering
\caption{Sample of test cases for Bug PL3}
\label{tab:test-pl3}

\resizebox{0.51\textwidth}{!}{
\setlength\tabcolsep{0.3 ex}
\begin{tabular}{|ccc|cc|c|}

\hline
\multicolumn{3}{|c|}{Input} & \multicolumn{2}{c|}{\scriptsize Output, \mycode{\scriptsize{abbreviate(str,lower,upper)}} } & \multirow{2}{*}{ \tabincell{c}{ Test \\ result} } \\ 
\cline{1-5}
 \mycode{str} & \mycode{\scriptsize lower} & \mycode{\scriptsize upper} &  Expected&  Observed& \\
\hline \hline
 \mycode{"0123456789"} & 0 & -1 & \mycode{"0123456789"} & \mycode{"0123456789"} & pass\\
 \mycode{"012 3456789"} & 0 &  5 & \mycode{"012"} & \mycode{"012 3456789"} & fail\\
\hline
\end{tabular}
}
\end{table}

A workaround would generate a runtime variable to collect existing side-effect free method calls with all possible parameters. For example, one could introduce a new variable \mycode{double tempVar = abs(x)} and generate a patch with the introduced variable \mycode{tempVar}. However, this workaround suffers from the problem of combinatorial explosion.

\subsubsection{Unavailable Method Values for a Null Object}
\label{subsubsect:limit-cl6}

Our repair approach can generate a patch with objected-oriented features. For example, a patch can contain 
state query methods on Java core library classes, such as \mycode{String.length()}, \mycode{File.exists()} and \mycode{Collection.size()}. We map these methods to their return values during the SMT encoding. However, such methods require that the object is not \mycode{null}; otherwise, a \mycode{null} pointer exception in Java is thrown. 

Let us consider Bug CL6, whose manually-written patch is \mycode{cs == null || cs.length() == 0}. For this bug, one passing test case detects whether the object \mycode{cs} is \mycode{null}. For this test case, the value of \mycode{cs.length()} is undefined and not given to SMT. Thus, it is impossible to generate a patch, which contains \mycode{cs.length()} if \mycode{cs} is \mycode{null} by at least one test case. Consequently, the SMT solver times-out because it tries to find a complex patch that satisfies the constraints. 

A possible solution is to encode the undefined values in the SMT. Constraints should be added to ensure that the unavailable values are not involved in the patch. This needs important changes in the design of the encoding, which is left to future work.

\section{Discussions}
\label{sect:discussions}

We now discuss \nopol features with respect to four important aspects.

\subsection{Differences with SemFix}
\label{subsect:discussion-semfix}

As mentioned in Section \ref{subsect:encoding}, \nopol uses the same technique in the phase of patch synthesis as SemFix \cite{nguyen2013semfix}, i.e., component-based program synthesis \cite{jha2010oracle}. However, there exist a number of important differences between \nopol and SemFix.

First, SemFix does not address missing pre-conditions. As shown in Table \ref{tab:patch}, adding preconditions enables us to repair more bugs than only updating conditions. We think that it is possible to extend the SemFix implementation to support repairing missing preconditions via adding the encoding strategy as in \nopol.

Second, \nopol does not use symbolic execution to find an angelic value. 
It is known that symbolic execution may have difficulties with the size and complexity of analyzed programs \cite{DBLP:journals/tosem/LiCZL14}. According to our work, we have the following observations. The angelic value in angelic fix localization is possible only when the domain is finite. For booleans, the domain of variables is not only finite but also very small. This results in a search space that can be explored dynamically and exhaustively as in \nopol. Symbolic execution as done in SemFix is capable of also reasoning on integer variables, because the underlying constraint solver is capable of exploring the integer search space. To sum up, our analytical answer is that for boolean domains, angelic fix localization is possible and probably much faster (this is claimed, but not empirically verified). For integer domains, only symbolic execution is appropriate. Meanwhile, \nopol can also handle the right-hand side of assignments as in SemFix. If we encode the synthesis as in SemFix, the right-hand side of assignments can be directly processed by \nopol.

Third, \nopol supports object-oriented code. We have adapted the code synthesis technique so that the generated patch can contain null checks and method calls. 

Finally, the evaluation of \nopol is significantly larger than that of SemFix. We have run \nopol on larger programs and real bugs. In SemFix, 4/5 of subject programs have less than 600 LoC and the bugs are artificially seeded. In the evaluation in our paper, the average number of lines of code per subject program is 25K LoC and the bugs are extracted from real programs that happened in practice.

\subsection{Effectiveness of Fault Localization Techniques}
\label{subsect:discussion-fault}

Fault localization plays an important role during the repair process. In our approach, a fault localization technique ranks all the suspicious statements and \nopol attempts to generate patches by starting with analyzing the most suspicious statement first. 
We use Ochiai \cite{abreu2007accuracy} as the fault localization technique. 
For our dataset, we wonder whether there is a difference between Ochiai and other techniques. In this section, we study the accuracy of different fault localization techniques on the bugs of our dataset.

We employ the absolute wasted effort to measure fault localization techniques. The wasted effort is defined as the ranking of the actual buggy statement. 
Given a set $S$ of statements, the wasted effort is expressed as follows, 

\vspace{-2ex}
$$effort=|\{susp(x)>susp(x^{*})\}|+1 $$
\vspace{-2ex}

\noindent where $x \in S$ is any statement, $x^{*}$ is the actual buggy statement, and $|\cdot|$ calculates the size of a set. A low value indicates that the fault localization technique is effective.

In our experiment, we compare six well-studied fault localization techniques: Ochiai \cite{abreu2007accuracy}, Tarantula \cite{jones2002visualization}, Jaccard \cite{abreu2007accuracy}, Naish2 \cite{naish2011model}, Ochiai2 \cite{naish2011model}, and Kulczynski2 \cite{xu2013general}. Table \ref{tab:localization} presents the comparison on the two types of bugs considered in this paper.

As shown in Table \ref{tab:localization}, for bugs with \buggyconditions, Jaccard obtains the best average wasted effort while Jaccard and Naish2 get the same median value. For bugs with \preconditions, Tarantula obtains the best average wasted effort while Ochiai, Jaccard, and Naish2 get the same median value. 
Those results assess that, according to our dataset of real bugs, the sensitivity of \nopol with respect to fault localization is not a crucial point and Ochiai is an acceptable choice.

\subsection{Potential Parallelization of \nopol}
\label{subsect:parallelization}

Our method \nopol is originally implemented to not perform parallelization. Based on the design of this method, it is possible to enhance the implementation by parallelizing \nopol to reduce the execution time. 
Indeed, the core algorithms of \nopol are highly parallelizable.

\begin{table}[!t]
\centering
\caption{Wasted effort comparison among six fault localization techniques.}
\label{tab:localization}
\setlength\tabcolsep{0.4 ex}
\resizebox{0.5\textwidth}{!}{
\begin{tabular}{|c|cc|cc|}

\hline

\multirow{2}{*}{ \tabincell{c}{Fault localization \\ technique}  }  & \multicolumn{2}{c|}{  Buggy \ourif condition}   & \multicolumn{2}{c|}{  Missing precondition} \\ \cline{2-5}
  & Average & Median  & Average & Median \\ 
\hline\hline
Ochiai  & 131.88  & 32.00 & 10.33  & \textbf{1.00} \\ 
Tarantula & 127.63  & 45.50 & \textbf{7.00}  & 1.50 \\ 
Jaccard & \textbf{121.44}  & \textbf{25.50} & 10.33  & \textbf{1.00} \\ 
Naish2  & 135.06  & \textbf{25.50} & 10.33  & \textbf{1.00} \\ 
Ochiai2 & 133.06  & 44.50 & 8.83  & 1.50 \\ 
Kulczynshi2 & 127.44  & 45.50 & 11.50  & 7.50 \\
\hline
\end{tabular}
}
\end{table}

First, during angelic fix localization, two feasible ways of parallelization can be performed: over test cases and over potential locations. 
Let us assume that there are 3 failing test cases with respectively 50, 100 and 200 \buggyconditions executed. Since the search space of \buggyconditions is $2\times n_c$ ($n_c$ is the number of executed \ourif statements by one test case, see Section \ref{subsubsect:search-space}), we could automatically run in parallel $(50+100+200)\times 2 = 700$ sessions of angelic fix localization on many different machines.

Second, our synthesis technique is based on different SMT levels (see Section \ref{sec:smt-level}). Synthesis at each SMT level corresponds to one independent SMT instance. Hence, the synthesis can also be run in parallel. However, we should mention that parallelizing the synthesis may lead to multiple resulted patches. Synthesis at a low SMT level can generate a simple patch; for the same bug, synthesis at a higher SMT level may not generate a better patch and may waste the running cost.

\subsection{Insight on Test-Suite Based Repair}
\label{subsect:discussion-repair}

\nopol is a test-suite based repair approach, as other existing work (\cite{le2012genprog,nguyen2013semfix,Kim2013,qi2014strength}, etc.) in the field of automatic software repair.
However, the foundations of test-suite based repair are little understood. Our experience with \nopol enables us to contribute to better understanding the strengths and the weaknesses of test-suite based repair.

There are two grand research questions behind test-suite based repair. The first one is about the quality of test suites \cite{monperrus2014critical}: do developers write good-enough test suites for automatic repair?
Qi et al. \cite{qi2015efficient} have shown that the test suites considered in the GenProg benchmark are not good enough, in the sense that they accept trivial repairs such as directly removing the faulty code. The experiments we have presented in this paper shed a different light. For nine bugs considered in this experiment, the test suite leads to a correct patch. For four additional bugs, a slight addition in the test suite allows for generating a correct patch. We consider this as encouraging for the future of the field. However, there is a need for future work on recommendation systems that tell when to add additional test cases for the sake of repair, and what those test cases should specify.

The second grand research question goes beyond standard test suites such as JUnit ones and asks whether repair operators do not overfit the inputs of input-output specifications \cite{Smith15fse}. 
For instance, for Bug CM6, one of the test inputs always equals to $-1$ when the buggy code is executed. As a result, the patch simply uses this value (corresponding to the number of rows) to drive the control flow, which is wrong. On the other hand, there are other cases when the repair operator yields a generic and correct solution upfront. This fact indicates that the same repair operator may overfit or not according to different bugs. It is necessary to conduct future research on the qualification of repair operators according to overfitting.

\section{Threats to Validity}
\label{sect:threats}

We discuss the threats to the validity of our results along four dimensions. 

\subsection{External Validity} 
\label{subsect:threat-number-bug}
In this work, we evaluate our approach on \numbug real-world bugs with \buggyconditions and \preconditions. One threat to our work is that the number of bugs is not large enough to represent the actual effectiveness of our technique. While the number of bugs in our work is fewer than that in previous work \cite{le2012genprog,nguyen2013semfix,Kim2013}, the main strength of our evaluation is twofold. On one hand, our work focuses on two specific types of bugs, i.e., \buggyandpres (as opposed to general types of bugs in \cite{Kim2013}); on the other hand, our work is evaluated on real-world bugs in large Java programs (as opposed to bugs in small-scale programs in \cite{nguyen2013semfix} and bugs without object-oriented features in \cite{le2012genprog}). We note that it is possible to collect more real-world bugs, with the price of more human labor. As mentioned in Section \ref{subsect:data-set}, reproducing a specific bug is complex and time-consuming. 

\subsection{Single Point of Repair} 
As all previous works in test-suite based repair, the program under repair must be repaired at one single point. In the current implementation of \nopol, we do not target programs with multiple faults, or bugs which require patches at multiple locations.

\subsection{Test Case Modification} 
In our work, we aim to repair bugs with \buggyconditions and \preconditions. Test cases are employed to validate the generated  patch.  In our experiment, several test cases are modified to facilitate repair. As mentioned in Section \ref{subsect:general-repair}, such test case modification consists of test case addition, transformation, and deletion. The answer to RQ\ref{rq-testcase} analyzes the root causes of test case modification. All test case modifications are listed in our project website \cite{nopol2014}.

\subsection{Dataset Construction} 
We describe how to construct our dataset in Section \ref{subsect:data-set}. The manually-written patches of conditional statements are extracted from commits in the version control system. However, it is common that a commit contains more code than the patch in \buggyandpres. In our work, we manually separate these patch fragments. In particular, the fixing commit of Bug PM2 contains two nested preconditions within a complex code snippet. We manually separate the patch of this bug according to the code context and keep only one precondition. 
Hence, there exists a potential bias in the dataset construction.

\section{Related Work}
\label{sect:relatedwork}

We list related work in four categories: approaches to test-suite based repair, repair besides test-suite based repair, empirical foundations of test-suite based repair, and related techniques in \nopol. 

\subsection{Test-Suite Based Repair}
\label{subsect:related-approach}

\textbf{GenProg}. Test-suite based repair generates and validates a patch with a given test suite. Le~Goues et al. \cite{le2012genprog} propose GenProg, an approach to test-suite based repair using genetic programming for C programs. In GenProg, a program is viewed as an Abstract Syntax Tree (AST) while a patch is a newly-generated AST by weighting statements in the program. Based on genetic programming, candidate patches are generated via multiple trials. 
The role of genetic programming is to obtain new ASTs by copying and replacing nodes in the original AST. 
A systematic study by Le~Goues et al. \cite{le2012systematic} shows that GenProg can fix 55 out of 105 bugs in C programs. 
The difference between \nopol and GenProg are as follows. 
\nopol targets a specific defect class while GenProg is generic;
\nopol uses component-based program synthesis while GenProg only copies existing code from the same code base;
\nopol uses a four-phase repair approach (fault localization, angelic fix localization, runtime trace collection, and patch synthesis) while GenProg uses a different two-phase approach (fault localization and trial);
\nopol is designed for object-oriented Java programs while GenProg is for C. 

\textbf{AE}. Weimer et al. \cite{DBLP:conf/kbse/WeimerFF13} report an adaptive repair method based on program equivalence, called AE. This method can fix 54 out of the same 105 bugs as in the work \cite{le2012systematic} while evaluating fewer test cases than GenProg. 

\textbf{PAR}. Kim et al. \cite{Kim2013} propose PAR, a repair approach using fix patterns representing common ways of fixing bugs in Java. These fix patterns can avoid nonsensical patches, which are caused by the randomness of some operators in genetic programming. Based on the fix patterns, 119 bugs are examined for patch generation. In this work, the evaluation of patches is contributed by 253 human subjects, including 89 students and 164 developers. 

\textbf{RSRepair}. Qi et al. \cite{qi2014strength} design RSRepair, a random search based technique for navigating the search space. This work indicates that random search performs more efficiently than genetic programming in GenProg \cite{le2012genprog}. RSRepair can fix 24 bugs, which are derived from a subset of 55 fixed bugs by GenProg \cite{le2012systematic}. 
Another work by Qi et al. \cite{DBLP:conf/icsm/QiML13} reduces the time cost of patch generation via test case prioritization. 

\textbf{SemFix}. Nguyen et al. \cite{nguyen2013semfix} propose SemFix, a constraint based repair approach. This approach generates patches for assignments and conditions by semantic analysis via SMT encoding. Program components are synthesized into one patch via translating the solution of the SMT instance. 
Our proposed approach, \nopol, is motivated by the design of SemFix. The major differences between \nopol and SemFix were discussed in Section \ref{subsect:discussion-semfix}. 

\textbf{Mutation-based repair}. Debroy \& Wong \cite{debroy2010using} develop a mutation-based repair method, which is inspired by the concept of mutation testing. Their method integrates program mutants with fault localization to explore the search space of patches. 

\textbf{DirectFix}. Mechtaev et al. \cite{mechtaev2015directfix} propose DirectFix, a repair method for simplifying patch generation. Potential program components in patches are encoded into a Maximum Satisfiability (MaxSAT) problem, i.e. an optimization problem; the solution to the MaxSAT instance is converted into the final concise patch.

\textbf{SearchRepair}. Ke et al. \cite{Ke15ase} develop SearchRepair, a repair method with semantic code search, which encodes human-written code fragments as SMT constraints on input-output behavior. This method reveals 20\% newly repaired defects, comparing with GenProg, AE, or RSRepair.

\textbf{SPR}. 
After the original publication presenting \nopol \cite{demarco2014automatic}, Long \& Rinard \cite{DBLP:conf/sigsoft/LongR15} have proposed a repair technique called SPR using condition synthesis. SPR addresses repairing conditional bugs, as well as other types of bugs, like missing non-\ourif statements. The differences are as follows.
First, a major difference between \nopol and SPR is that \nopol synthesizes a condition via component-based program synthesis while SPR is based on multiple trials of pre-defined program transformation schemas. For instance, in SPR, a transformation schema for conditional bugs is called \textit{condition refinement}, which updates an existing condition in an \ourif via tightening or loosening the condition. To repair a bug, SPR tries a potential patch with the transformation schemas one by one and validates the patch with the test suite; the technique of \nopol is entirely different, based on runtime data collection during test execution. 
Second, another difference is that SPR is for repairing C programs. Patches by SPR only contain primitive values while patches by \nopol contain both primitive values and object-oriented expressions (e.g., fields and unary method calls). 
Third, in SPR, the technique of collecting angelic values is based on \nopol's, yet extends it.
It finds sequences of values rather than one simplified trace during collecting angelic values in \nopol (Section~\ref{sect:angelic-fix-localization}). As mentioned in Section \ref{subsubsect:search-space}, the simplified trace in \nopol reduces the search space of patch synthesis, but may result in failed repair attempts for specific bugs, where a condition is executed more than once by a test case. Examples of these bugs can be found in the SPR evaluation \cite{DBLP:conf/sigsoft/LongR15}. The simplification in \nopol can be viewed as a trade-off between repairability and time cost. 

\textbf{Prophet}. Also by Long \& Rinard, Prophet is an extension of SPR that uses a probability model for prioritizing candidate patches. Based on historical patches, Prophet learns model parameters via maximum likelihood estimation. Experiments show that this method can generate correct patches for 15 out of 69 real-world defects of the GenProg benchmark. We have also noticed that in \nopol, it is possible to synthesize more than one patch with our SMT-based synthesis implementation. Hence, the probability model in Prophet can be leveraged to direct the synthesis of more correct patches by \nopol.  

\subsection{Other Kinds of Repair}
\label{subsect:other-kinf-repair}

Besides test-suite based repair, other approaches are designed for fixing software bugs and improving software quality. 
Dallmeier et al. \cite{dallmeier2009generating} propose Pachika, a fix generation approach via object behavior anomaly detection. This approach identifies the difference between program behaviors by the execution of passing and failing test cases; then fixes are generated by inserting or deleting method calls. 
Carzaniga et al. \cite{carzaniga2010automatic} develop an automatic technique to avoid failures by a faulty web application. This technique is referred as an automatic workaround, which aims to find and execute a correct program variant. 
AutoFix by Pei et al. \cite{pei2014automated}, employs a contract-based strategy to generate fixes. This approach requires simple specifications in contracts, e.g., pre-conditions and post-conditions of a function, to enhance the debugging and fixing process. Experiments on Eiffel programs show that this approach can fix 42\% of over 200 faults.

\subsection{Empirical Foundations of Repair}
\label{subsect:related-foundation}

Applying automatic repair to real-world programs is limited by complex program structures and semantics. We list existing work on the investigation of empirical foundations of test-suite based repair. 

Martinez \& Monperrus \cite{Martinez2013} mine historical repair actions to reason about future actions with a probabilistic model. Based on a fine granularity of ASTs, this work analyzes over 62 thousands versioning transactions in 14 repositories of open-source Java projects to collect probabilistic distributions of repair actions. Such distributions can be used as prior knowledge to guide program repairing. 

Fry et al. \cite{DBLP:conf/issta/FryLW12} design a human study of patch maintainability with 150 participants and 32 real-world defects. This work indicates that machine-generated patches are slightly less maintainable than human-written ones; hence, patches by automatic repair could be used as the patches written by humans. Another case study is conducted by Tao et al. \cite{DBLP:conf/sigsoft/TaoKKX14}. They investigate the possibility of leveraging patches by automatic repair to assist the process of debugging by humans.  

Barr et al. \cite{DBLP:conf/sigsoft/BarrBDHS14} address the ``plastic surgery hypothesis'' of genetic-programming based repair, such as GenProg. Their work presents evidences of patches based on reusable code, which make patch reconstitution from existing code possible. Martinez et al. \cite{DBLP:conf/icse/MartinezWM14} conduct empirical investigation to the redundancy assumption of automatic repair; this work indicates that code extracted from buggy programs could form a patch that passes the test suite.  

Monperrus \cite{monperrus2014critical} details the problem statement and the evaluation of automatic software repair. This work systematically describes the pitfalls in software repair research and the importance of explicit defect classes; meanwhile, this paper identifies the evaluation criteria in the field: understandability, correctness, and completeness. 
Zhong \& Su \cite{zhong2015an} examine over 9,000 real-world patches and summarize 15 findings in two key ingredients of automatic repair: fault localization and faulty code fix. This work provides empirical foundations for localization and patch generation of buggy statements.

Qi et al. \cite{qi2015efficient} propose Kali, an efficient repair approach based on simple actions, such as statement removal. Their work presents the repair results via simple methods; meanwhile, their work checks previous empirical results by GenProg \cite{le2012genprog}, AE \cite{DBLP:conf/kbse/WeimerFF13}, and RSRepair \cite{DBLP:conf/icsm/QiML13}. Empirical studies show that only two bugs by GenProg, three bugs by AE, and two bugs by RSRepair are correctly patched. All the reported patches for the other bugs are incorrect due to improper experimental configurations or semantic issues; an incorrect patch either fails to produce expected outputs for the inputs in the test suite, or fails to implement functionality that is expected by developers. 
As the latest result in test-suite based repair, the work by Qi et al. \cite{qi2015efficient} shows that repairing real-world bugs is complex and difficult. Hence, it is worth investigating the empirical results on fixing real bugs. 

Recent work by Smith et al. \cite{Smith15fse} investigates the overfitting patches on test cases in automatic repair. They report a controlled experiment on a set of programs written by novice developers with bugs and patches; two typical repair methods, GenProg \cite{le2012genprog} and RSRepair \cite{qi2014strength}, are evaluated to explore the factors that affect the output quality of automatic repair. 

Recent work by Le Goues et al. \cite{LeGoues15tse} presents two datasets of bugs in C programs to support comparative evaluation of automatic repair algorithms. The detailed description of these datasets is introduced and a quantified empirical study is conducted on the datasets. 
Defects4J by Just et al. \cite{JustJE2014} is a bug database that consists of 357 real-world bugs from five widely-used open-source Java projects. It has recently been shown \cite{durieux2015} that \nopol is capable of fixing 35 bugs of this benchmark.

\subsection{Related Techniques: Program Synthesis and Fault Localization}
\label{subsect:related-localization}

Our approach, \nopol, relies on two important techniques, program synthesis and fault localization. 

Program synthesis aims to form a new program by synthesizing existing program components. Jha et al. \cite{jha2010oracle} mine program oracles based on examples and employ SMT solvers to synthesize constraints. In this work, manual or formal specifications are replaced by input-output oracles. They evaluate this work on 25 benchmark examples in program deobfuscation. Their follow-up work \cite{gulwani2011synthesis} addresses the same problem by encoding the synthesis constraint with a first-order logic formula. In general, any advance in program synthesis can benefit program repair by enabling either more complex or bigger expressions to be synthesized. 

In our work, fault localization is used as a step of ranking suspicious statements to find out locations of bugs. A general framework of fault localization is to collect program spectra (a matrix of testing results based on a given test suite) and to sort statements in the spectra with specific metrics (e.g., Tarantula \cite{jones2002visualization} and Ochiai \cite{abreu2007accuracy}). Among existing metrics in fault localization, Ochiai \cite{abreu2007accuracy} has been evaluated as one of the most effective ones. In Ochiai, statements are ranked according to their suspiciousness scores, which are values of the Ochiai index between the number of failed test cases and the number of covered test cases. Fault localization techniques are further improved recently, for example, the diagnosis model by Naish et al. \cite{naish2011model}, the localization prioritization by Yoo et al.~\cite{yoo2013fault}, and the test purification by Xuan \& Monperrus~\cite{xuan2014test}.

\section{Conclusion}
\label{sect:conclusion}

In this paper, we have proposed \nopol, a test-suite based repair approach using SMT. \nopol targets two kinds of bugs: buggy \ourif conditions and missing preconditions. Given a buggy program and its test suite, \nopol employs angelic fix localization to identify potential locations of patches and expected values of \ourif conditions. For each identified location, \nopol collects test execution traces of the program. Those traces are then encoded as an SMT problem and the solution to this SMT is converted into a patch for the buggy program. We conduct an empirical evaluation on \numbug real-world programs with \buggyconditions and \preconditions. We have presented four case studies to show the benefits of generating patches with \nopol as well as the limitations. 

\nopol is publicly-available to support further replication and research on automatic software repair: \url{http://github.com/SpoonLabs/nopol/}.

In future work, we plan to evaluate our approach on more real-world bugs. Our future work also includes addressing the current limitations, e.g., designing better strategy for angelic fix localization, collecting more method calls, and improving the SMT encoding.

\section*{Acknowledgment}

The authors would like to thank David Cok for giving us full access to jSMTLIB. 
This work is partly supported by the INRIA Internship program, the INRIA postdoctoral research fellowship, the CNRS delegation program, the National Natural Science Foundation of China (under grant 61502345), and the Young Talent Development Program of the China Computer Federation.

\balance
\bibliographystyle{abbrv}
\bibliography{references}

\begin{IEEEbiography} [{\includegraphics[width=1in,height=1.25in,clip,keepaspectratio]{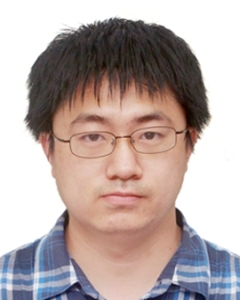}}]
{Jifeng Xuan} received the BSc degree in software engineering in 2007 and the PhD degree in 2013, both from Dalian University of Technology, China. He is a research professor at the State Key Lab of Software Engineering, Wuhan University, China. He was previously a postdoctoral researcher at the INRIA Lille -- Nord Europe, France. His research interests include software testing and debugging, software data analysis, and search based software engineering. 
He is a member of the IEEE. 
\end{IEEEbiography}

\begin{IEEEbiography} [{\includegraphics[width=1in,height=1.25in,clip,keepaspectratio]{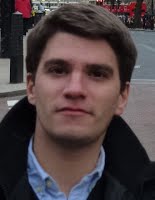}}]
{Matias Martinez} received the Master degree in computer engineering from UNICEN, Argentina, and the PhD degree for University of Lille, France, in 2014. He is currently a postdoctoral researcher at University of Lugano (USI), Switzerland. His current research focuses on automatic software repair, software testing, and mining software repositories. 
\end{IEEEbiography}

\begin{IEEEbiography} [{\includegraphics[width=1in,height=1.25in,clip,keepaspectratio]{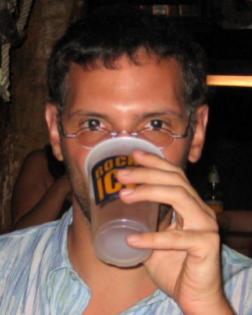}}]
{Favio DeMarco} received the MSc degree in Computer Science from Universidad de Buenos Aires, Argentina. His research interests focus on software repair. 
\end{IEEEbiography}

\begin{IEEEbiography} [{\includegraphics[width=1in,height=1.25in,clip,keepaspectratio]{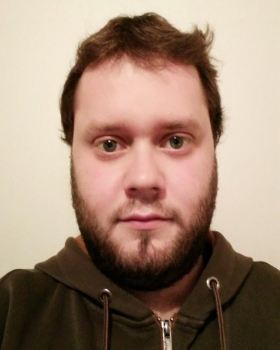}}]
{Maxime Cl\'{e}ment} is the MSc degree student in computer engineering from Lille 1 University, France. He is currently working in an IT services company as part of an internship program.
\end{IEEEbiography}

\begin{IEEEbiography} [{\includegraphics[width=1in,height=1.25in,clip,keepaspectratio]{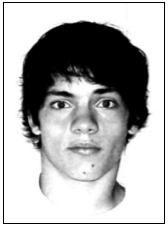}}]
{Sebastian Lamelas Marcote} received the MSc degree in Computer Science from Universidad de Buenos Aires (FCEN), Argentina in 2015. He is currently working in the industry, but still has a research spirit focused on the integration of computer power with other scientific disciplines. 
\end{IEEEbiography}

\begin{IEEEbiography} [{\includegraphics[width=1in,height=1.25in,clip,keepaspectratio]{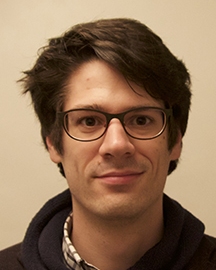}}]
{Thomas Durieux} received the master degree in computer engineering from University of Lille, France in 2015. He is working toward the PhD degree in software engineering at the University of Lille. His current research focuses on automatic repair at runtime.
\end{IEEEbiography}

\begin{IEEEbiography} [{\includegraphics[width=1in,height=1.25in,clip,keepaspectratio]{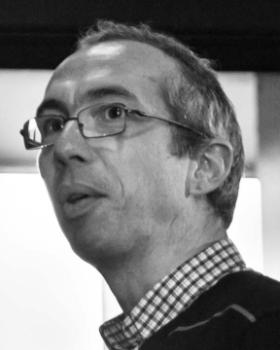}}]
{Daniel Le Berre} is a professor of computer science at Artois University, Lens, France, and a member of the CRIL research lab, a research center specialized in artificial intelligence.
His main research interests lie around Boolean constraint programming and its applications to software engineering. 
\end{IEEEbiography}

\begin{IEEEbiography} [{\includegraphics[width=1in,height=1.25in,clip,keepaspectratio]{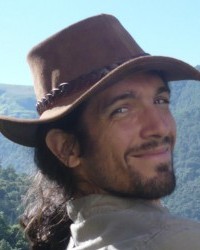}}]
{Martin Monperrus} has been an associate professor at the University of Lille, France, since 2011. He is an adjunct researcher at Inria. He was previously with the Darmstadt University of Technology, Germany, as a research associate. He received a Ph.D. from the University of Rennes in 2008. His research lies in the field of software engineering with a current focus on automatic software repair. He is a member of the IEEE.
\end{IEEEbiography}

\end{document}